\documentclass[journal]{IEEEtran}
\usepackage{amsmath,amssymb,verbatim,cite,amsopn,graphicx,url,color}
\usepackage{algorithm}
\usepackage{algorithmic}
\usepackage{multicol}
\usepackage{epsfig}
\usepackage{epstopdf}
\usepackage{subcaption}
\usepackage{balance}
\usepackage[font={small}]{caption}
\usepackage{bm}
\def\tablename{Table}
\renewcommand{\thetable}{\arabic{table}}

\newcommand{\LB}{\text{LB}}
\newcommand{\UB}{\text{UB}}

\newcommand{\SINR}{\text{SINR}}

\newcommand{\psok}{p_{\text{so},k}}
\newcommand{\psopik}{p_{\text{so},\pi(k)}}
\newcommand{\Rsmin}{R_{s,\min}}

\newcommand{\TDMA}{\text{TDMA}}
\newcommand{\NOMA}{\text{NOMA}}

\newcommand{\up}{\text{up}}

\newtheorem{Proposition}{Proposition}
\newtheorem{Theorem}{Theorem}

\newtheorem{Corollary}{Corollary}
\newtheorem{Remark}{Remark}

\begin{document}
\title{On the Design of Secure Non-Orthogonal Multiple Access Systems\vspace{4mm}}

\author{
Biao He,~\IEEEmembership{Member,~IEEE,}
An Liu,~\IEEEmembership{Member,~IEEE,}
Nan Yang,~\IEEEmembership{Member,~IEEE,}
and Vincent K. N. Lau,~\IEEEmembership{Fellow,~IEEE}
\vspace{-5mm}
\thanks{B. He is with the Center for Pervasive Communications and Computing,
University of California at Irvine, Irvine, CA 92697, USA (email: biao.he@uci.edu).}
\thanks{A. Liu, and V. K. N. Lau are with the Department of Electronic and Computer Engineering, The Hong Kong University of Science and Technology, Hong Kong (email:
\{eewendaol, eeknlau\}@ust.hk).}
\thanks{N. Yang is with the Research School of Engineering, The Australian National University, Canberra, ACT 2601, Australia (e-mail: nan.yang@anu.edu.au).}
}

\maketitle

\begin{abstract}
This paper proposes a new design of non-orthogonal multiple access (NOMA) under secrecy considerations.
We focus on a NOMA system where a transmitter sends confidential messages to multiple users in the presence of an external eavesdropper.
The optimal designs of decoding order, transmission rates, and power allocated to each user are investigated.
Considering the practical passive eavesdropping scenario where the instantaneous channel state of the eavesdropper is unknown, we adopt the secrecy outage probability as the secrecy metric.
We first consider the problem of minimizing the transmit power subject to the secrecy outage and quality of service constraints, and derive the closed-form solution to this problem. We then explore the problem of maximizing the minimum confidential information rate among users subject to the secrecy outage and transmit power constraints, and provide an iterative algorithm to solve this problem. We find that the secrecy outage constraint in the studied problems does not change the optimal decoding order for NOMA, and one should increase the power allocated to the user whose channel is relatively bad when the secrecy constraint becomes more stringent. Finally, we show the advantage of NOMA over orthogonal multiple access in the studied problems both analytically and numerically.
\end{abstract}


\section{Introduction}\label{sec:Intro}
Non-orthogonal multiple access (NOMA) is envisaged as a  potentially promising technique to address some key challenges in the fifth generation (5G) networks, such as high spectral efficiency and massive connectivity~\cite{Saito_13_systemnoma,Dai_16_NOMAsco_mag}.
Different from the traditional orthogonal multiple access (OMA) techniques that rely on the orthogonal resource allocation, such as time-division multiple access (TDMA),
the NOMA explores the non-orthogonal resource allocation.
Two dominant NOMA solutions for future networks have been proposed, which are the power domain multiplexing and the code domain multiplexing~\cite{Dai_16_NOMAsco_mag}.
For the power domain multiplexing, different users are allocated with different power
levels according to their channel conditions, and the successive interference cancellation (SIC) is used to cancel multi-user interference. For the code domain multiplexing, e.g., sparse code multiple access, different users are assigned with different codes, and then multiplexed over the same time-frequency resources.
The performance of NOMA in the network with randomly deployed users was analyzed in~\cite{Ding_14_ontpnmai}. The design problem of the NOMA scheme from a fairness standpoint was addressed in~\cite{Timotheou_15_FairnessNOMA5G}.
Cooperative NOMA schemes were investigated in \cite{NOMADCTPS_14_Choi} and \cite{Ding_14_CoopNonMA5S} to improve the system reliability. To provide additional spatial degrees of freedom, multiple-input multiple-output (MIMO) techniques were introduced to NOMA  systems in \cite{Choi_15_MinPMBSCM}  and~\cite{Ding_16_MIMONOMAframework}.
The hybrid multiple access system combining NOMA and traditional OMA by the technique of user pairing was studied in~\cite{Ding_16_IuserpairingNMADT}.

As ubiquitous wireless devices are adopted in modern life, an unprecedented amount of private and sensitive data is transmitted over wireless channels. Consequently, the secrecy issues associated with wireless networks have become critical. On the other hand, the information exchange between transceivers over wireless channels is vulnerable to eavesdropping, due to the unalterable open nature of wireless medium. Therefore, the research on the secrecy issue of wireless transmissions is of significant importance.
Physical layer security has been widely regarded as a promising complement to the
cryptographic techniques to secure the data transmission over wireless channels~\cite{Zhou_13_Physical,Yang_15_Mag}.
In the seminal work~\cite{wyner_75}, Wyner introduced the wiretap channel as a framework for physical layer security, where a transmitter wants to send confidential messages to a legitimate receiver in the presence of an external eavesdropper. Following Wyner's wiretap channel, the wireless physical layer security against external eavesdroppers has been extensively studied in recent years, e.g., \cite{Gopala_08,Liang_08,Bloch_08} from an information-theoretic perspective on the performance analysis and \cite{Goel_08,Mukherjee_11,He_13_2} from a signal-processing perspective on the system design.
It is worth mentioning that there is another research direction on physical layer security, which does not focus on the secrecy issue against external eavesdroppers. Instead, confidential messages to the intended users are required to remain ignorance at other users, e.g.,~\cite{csiszar_78,Liu_10_Multiple,Geraci_13_large,Yang_14_Confidential,He_15_Base}.

Although a great amount of research effort has been paid to physical layer security in wireless communications, the secrecy issue of NOMA has been rarely considered in the literature, except for the recent studies~\cite{Qin_16_ICC_PLSNOMA,Zhang_16_ssrNOMA,Ding_16_OnSEandEENOMAMUC}.
On the other hand, in future wireless networks, it is indeed pivotal to explore the security level of the non-orthogonal transmission among multiple users.
The secrecy issues of NOMA against external eavesdroppers were studied in~\cite{Qin_16_ICC_PLSNOMA} and \cite{Zhang_16_ssrNOMA}, where the messages sent to all users by NOMA are required to be secure against the external eavesdropper.
Considering large-scale networks, the secrecy performance of NOMA with randomly deployed users and eavesdroppers were analyzed in~\cite{Qin_16_ICC_PLSNOMA} by using stochastic geometry~\cite{Haenggi_12}. Focusing on the design of NOMA scheme, the problem of maximizing the secrecy sum rate at all users subject to a qualify of service (QoS) constraint of the codeword rate at each user was tackled in~\cite{Zhang_16_ssrNOMA}.
Note that the perfect secrecy rate/capacity considered in~\cite{Zhang_16_ssrNOMA} can be ensured only when the transmitter has the perfect knowledge on the eavesdropper's channel state information (CSI), which is difficult to realize in practice.
In addition, the QoS constraint of the codeword rate cannot guarantee the rate of informative data received by the user, although it simplifies the design problem.
Instead of considering the external eavesdropper, \cite{Ding_16_OnSEandEENOMAMUC} investigated the NOMA system where the transmitter wants to send the confidential message to only one user and keep it ignorance at all other users.

From the above discussion, we note that the design problem of NOMA schemes against the external eavesdropper with appropriate secrecy and QoS performance metrics has yet been investigated. This motivates us to design the secure NOMA schemes for the practical scenario where the transmitter does not know the eavesdropper's instantaneous channel information.
In this scenario, the perfect secrecy rate is usually not achievable, and hence, we do not take it as the secrecy  metric. Instead, we use the secrecy outage probability to measure the secrecy performance of the system~\cite{Zhou_11}.  We highlight that the secrecy outage probability is an appropriate secrecy metric for the systems where the eavesdropper's CSI is not perfectly known~\cite{Bloch_08,He_13_2,He_16_OnMetrics}.
The primary contributions of the paper are summarized as follows.
\begin{itemize}
\item We comprehensively investigate the design of NOMA against the external eavesdropper under the secrecy outage constraint. 
    The decoding order, transmission rates, and power allocated to each user are all considered as designable parameters.
    For the first time, we analytically prove that the optimal decoding order for the NOMA scheme with the secrecy outage constraint is the same as that for the conventional NOMA scheme without the secrecy constraint, i.e., the descending order of channel gains normalized by noise \cite{Tse_05_Fundamentals}.
\item We derive the closed-form solution to the problem of minimizing the transmit power subject to the secrecy outage constraint and a QoS constraint.
    Different from \cite{Zhang_16_ssrNOMA}, the QoS performance is characterized by the confidential information rate at each user, which appropriately captures the real rate of informative data received at the user.
    The problem is non-convex. Unlike many conventional design problems of NOMA, e.g.,~\cite{Timotheou_15_FairnessNOMA5G,Zhang_16_ssrNOMA,Cui_16_ANPASUOCIS}, our problem cannot be transformed into a sequence of liner programs (LPs) due to the secrecy outage constraint, and hence, it is more difficult to solve.
\item We further solve the problem of maximizing the minimum confidential information rate among users subject to the secrecy outage constraint and an instantaneous transmit power constraint. Here, we regard the minimum confidential information rate among users as a fairness performance of NOMA with the secrecy consideration, which has never been investigated in the literature. We propose an iterative algorithm to solve the problem, and obtain the closed-form solution to the problem for the special case of two users. 
    We find that one should increase the ratio of power allocated to the user who has a relatively bad channel when the secrecy constraint becomes more stringent.
\item Both analytically and numerically, we compare the performance of the NOMA scheme with that of the OMA scheme in the studied problems with the secrecy outage constraint.  We analytically prove that the NOMA scheme always outperforms the OMA scheme, and numerically show that the performance gain of the NOMA scheme over the OMA scheme increases almost linearly as the number of users increases.
\end{itemize}

The remainder of the paper is organized as follows. Section~\ref{sec:sysmod} presents the system model. Sections~\ref{sec:mimpsum} and~\ref{sec:maxFairnessRate} study the problem of minimizing the transmit power subject to the secrecy outage and QoS constraints and the problem of maximizing the minimum confidential information rate subject to the secrecy outage and transmit power constraints, respectively.
The analytical comparison between NOMA and OMA in the studied problems is conducted in Section~\ref{sec:analyticalcomp}.
The numerical results are presented in Section~\ref{Sec:NumericalResults}.
Finally, Section~\ref{sec:conclus} concludes the paper.

\section{System Model}\label{sec:sysmod}
\begin{figure}[!t]
\centering
\includegraphics[width=0.9\columnwidth]{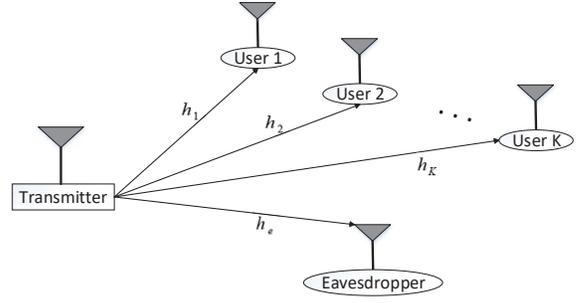}
\caption{Illustration of a NOMA system in the presence of an external eavesdropper.}
 \label{fig:SystemModel}
\end{figure}

As shown in Figure~\ref{fig:SystemModel}, we consider a communication system where a transmitter wants to send confidential messages to $K$ users in the presence of an eavesdropper.
We assume that the eavesdropper is external, and the transmitter can distinguish the eavesdropper from its own users. In practice, the transmitter can identify its users through the authentication process before the data transmission.
The transmitter, the users, and the eavesdropper each have a  single antenna. For the sake of brevity, we denote the $k$-th user as user $k$, $k\in\left\{1,\cdots,K\right\}$.
The channel coefficients from the transmitter to  user $k$ and the eavesdropper are denoted by $h_k$ and $h_e$, respectively. Both path loss and fading effects are considered, such that $h_k=d_k^{-\alpha/2}g_k$ and $h_e=d_e^{-\alpha/2}g_e$, where $\alpha$ denotes the path-loss exponent, $d_k$ and $d_e$ denote the distances from the transmitter to user $k$ and the eavesdropper, respectively, and $g_k,~g_e\sim\mathcal{CN}\left(0,1\right)$ denote the normalized Rayleigh fading channel states. The quasi-static block fading model is adopted~\cite{Ozarow_94}, such that the channel coefficients remain constant during the transmission of one message, which consists of a block of symbols, and change independently from one block to the next. We  assume that all channels are independent of each other. Without loss of generality, the channel gains are sorted as $0<\left|h_1\right|^2\le\cdots\le\left|h_K\right|^2$.
We further assume that the transmitter knows the instantaneous channel gains of all users, i.e.,  $\left|h_k\right|^2$, but knows only  the average channel gain of the eavesdropper  over different fading realizations, i.e,  $\mathbb{E}\left\{\left|h_k\right|^2\right\}$, where $\mathbb{E}\{\cdot\}$ denotes the expectation operation.
Note that $\mathbb{E}\left\{\left|h_k\right|^2\right\}=d_e^{-\alpha}$ can be obtained at the transmitter from the knowledge of the distance between the transmitter and the eavesdropper.
Here, we emphasize that the transmitter does not have the instantaneous knowledge on the eavesdropper's channel.
It is worth mentioning that even the statistics of the eavesdropper's channel may not be easy to obtain. Ideally, the physical layer security can be achieved without knowing any CSI of the eavesdropper. Unfortunately, this is not achievable by the physical layer security techniques in most wireless networks. Therefore, we have used the widely-adopted assumption of knowing the statistics of the eavesdropper's channel; see, e.g., \cite{Bloch_08,Zhou_13_Physical,He_16_OnMetrics} and references within.
In practice, the channel statistics of the eavesdropper can be estimated per the knowledge of the fading environment and the distance of the eavesdropper. For example, the Rayleigh fading model is often assumed for the rich-scattering environment with  no line-of-sight propagation between the transceivers, and the average channel gain can be estimated based on the distance between the transceivers.
\vspace{-4mm}

\subsection{NOMA Scheme}
The NOMA scheme enables the transmitter to simultaneously serve multiple users~\cite{Saito_13_systemnoma,Ding_14_ontpnmai}. The transmitter uses the superposition coding (SC) to send a linear combination of multiple signals to the users. 
The transmitted signal is given by $\sum_{k=1}^K \sqrt{P_k}s_k$, where $P_k$ denotes the transmit power allocated to user $k$  and $s_k$ denotes the normalized message for user $k$.
The instantaneous total transmit power is $\sum_{k=1}^K P_k$.
The received signal at user $k$ and the eavesdropper are, respectively, given by
\begin{equation}\label{}
  y_k=h_k\sum_{k=1}^K \sqrt{P_k}s_k+n_k
\end{equation}
and
\begin{equation}\label{}
  y_e=h_e\sum_{k=1}^K \sqrt{P_k}s_k+n_e,
\end{equation}
where $n_k$ and $n_e$ denote the zero-mean additive white Gaussian noise (AWGN) at user $k$ with variance $\sigma_k^2$ and the zero-mean AWGN at the eavesdropper with variance $\sigma_e^2$, respectively. We assume that the noise variances at all users are identical, i.e., $\sigma_1^2=\cdots=\sigma_k^2\!=\!\sigma_u^2$.

As per the mechanism of NOMA, all users adopt the SIC to decode messages with the same decoding order~\cite{Tse_05_Fundamentals}. Note that for NOMA systems with the secrecy consideration, it is not clear what the optimal decoding order is.
Thus, the $m$-th message to be decoded at users may not be the message for user $m$.
As such, we need to introduce the variable $\pi$ to denote the decoding order and optimize the decoding order $\pi$ as well.
For example, if $\pi(1)=3$, then the first message to be decoded for the SIC is the message for user 3. Specifically, user $\pi(k)$ decodes the messages for all users $\pi(j)$, $\forall~j<k$, before decoding its own message to remove the inter-user interference in a successive way.
Then, user $\pi(k)$ decodes its own message while treating the messages for all users $\pi(i)$, $\forall~i>k$,  as the interference.
The received signal-to-interference-plus-noise ratios (SINRs) at user $\pi(k)$, $k<K$, and user $\pi(K)$ to decode their own messages are given by 
\begin{align}\label{eq:SINRkk}
  \SINR_{\pi(k)} &=\frac{\left|h_{\pi(k)}\right|^2P_{\pi(k)}}{\left|h_{\pi(k)}\right|^2\sum_{i=k+1}^{K}P_{\pi(i)}+\sigma_u^2} \notag\\
  &=\frac{\gamma_{\pi(k)}P_{\pi(k)}}{1+\gamma_{\pi(k)}\sum_{i=k+1}^{K}P_{\pi(i)}},~~~~k<K
\end{align}
and
\begin{equation}\label{}
  \SINR_{\pi(K)} =\frac{\left|h_{\pi(K)}\right|^2P_{\pi(K)}}{\sigma_u^2}=\gamma_{\pi(K)}P_{\pi(K)},
\end{equation}
respectively, where
$\gamma_{\pi(k)}=\left|h_{\pi(k)}\right|^2/\sigma_u^2$.
In addition, the received SINR at user ${\pi(m)}$ to decode the message $s_{\pi(k)}$, $k<m\le K$, is given by
\begin{align}\label{eq:SINRkj}
  \SINR_{{\pi(m)},{\pi(k)}} &=\frac{\left|h_{\pi(m)}\right|^2P_{\pi(k)}}{\left|h_{\pi(m)}\right|^2\sum_{i=k+1}^{K}P_{\pi(i)}+\sigma_u^2}
  \notag\\
  &=\frac{\gamma_{\pi(m)}P_{\pi(k)}}{1+\gamma_{\pi(m)}\sum_{i=k+1}^{K}P_{\pi(i)}},~~~~k<m\le K.
\end{align}
Similarly, the received SINRs at the eavesdropper of the message $s_k$, $k<K$, and the message $s_K$ are given by~\cite{Zhang_16_ssrNOMA}
\begin{align}\label{eq:eavesSINRk}
  \SINR_{\tilde{\pi}(k)}&=\frac{\left|h_e\right|^2P_{\pi(k)}}{\left|h_e\right|^2\sum_{i=k+1}^{K}P_{\pi(i)}+\sigma_e^2}
\notag\\
&=\frac{\gamma_eP_{\pi(k)}}{1+\gamma_e\sum_{i=k+1}^{K}P_{\pi(i)}},~~~~k<K
\end{align}
and
\begin{equation}\label{eq:eavesSINRK}
  \SINR_{\tilde{\pi}(K)}=\frac{\left|h_e\right|^2P_{\pi(K)}}{\sigma_e^2}=\gamma_eP_{\pi(K)},
\end{equation}
respectively, where
$\gamma_e=\left|h_e\right|^2/\sigma_e^2$.
Note that the expressions for the received SINRs at eavesdropper here overestimate the eavesdropper's capability.
A worst-case assumption from the legitimate users' perspective is made here.
That is, the eavesdropper has already decoded the messages for all users ${\pi(j)}$, $\forall~j<k$, before it attempts to decode the message for user ${\pi(k)}$.
This assumption also implies that the eavesdropper knows the decoding order and the power allocation.
In fact, the eavesdropper may or may not know the users' decoding order and the power allocation, and may or may not know the messages for all users ${\pi(j)}$, $\forall~j<k$, before it attempts to decode the message for user ${\pi(k)}$. However, the legitimate users cannot know the eavesdropper's knowledge, since the eavesdropper would not inform the legitimate users about its ability and the instantaneous CSI. Thus, we have to adopt the worst-case assumption from the legitimate users' perspective due to the conservativeness mandated by the security studies.
Note that this assumption has been adopted in the previous work on the secrecy of NOMA systems \cite{Zhang_16_ssrNOMA}. In addition, we highlight that worst-case assumptions have been widely adopted in analyzing and designing transmission schemes with secrecy considerations; see, e.g.,~\cite{Huang_12,He_15_Base,He_15_Achieving}.

It has been clearly shown that the capacity-achieving decoding order for conventional NOMA schemes is in the descending order of channel gains normalized by noise, i.e., channel strengths \cite{Tse_05_Fundamentals,Wei_16_ASurveyNOMAZTE}.
However, to the best of the authors' knowledge, the optimal decoding order for NOMA schemes with the secrecy consideration has never been analyzed or discussed in the existing literature.

It is worth mentioning that the considered NOMA scheme in this paper is the power domain NOMA, and the proposed secrecy scheme is applicable to systems with power domain NOMA. The proposed scheme is not applicable to the code domain NOMA, e.g., sparse code multiple access, where different users are assigned with different codes multiplexed over the same time-frequency resources~\cite{Dai_16_NOMAsco_mag}.

\subsection{Secure Encoding}
We consider the widely-adopted wiretap code \cite{wyner_75} to secure the transmission of messages. There are two rate parameters for the message to each user $k$, namely, the codeword transmission rate, $R_{t,k}$, and the confidential information rate, $R_{s,k}$. The positive rate difference $R_{t,k}-R_{s,k}$ is the rate cost to provide secrecy against the eavesdropper.
A length $n$ wiretap code is constructed by generating $2^{nR_{t,k}}$ codewords $x^n(w,v)$, where $w=1,\cdots,2^{nR_{s,k}}$ and $v=1,\cdots,2^{n\left(R_{t,k}-R_{s,k}\right)}$.
For each message index $w$, we randomly select $v$ from $\left\{1,2,\cdots,2^{n\left(R_{t,k}-R_{s,k}\right)}\right\}$ with uniform probability and transmit the codeword $x^n(w,v)$.
We consider the adaptive-rate transmission, where the rate parameters $R_{t,k}$ and $R_{s,k}$ can be adaptively chosen according to the instantaneous CSI of the users.

Since the quasi-static fading channel is considered and the eavesdropper's instantaneous CSI is unknown at the transmitter, perfect secrecy is not achievable. The secrecy outage probability is adopted to measure the secrecy performance of the transmission~\cite{Bloch_08,Zhou_11,He_16_OnMetrics}. The secrecy outage probability of the message $s_{\pi(k)}$ is given by~\cite{Zhou_11}
\begin{equation}\label{eq:psokdefination}
  \psopik=\mathbb{P}\left(R_{t,\pi(k)}-R_{s,\pi(k)}<C_{e,\pi(k)}\right),
\end{equation}
where $\mathbb{P}(\cdot)$ denotes the probability measure and $C_{e,\pi(k)}=\log_2\left(1+SNR_{\tilde{\pi}(k)}\right)$ denotes the eavesdropper's channel capacity to decode the message $s_{\pi(k)}$.

\section{Transmit Power Minimization}\label{sec:mimpsum}
In this section, we design the NOMA scheme that minimizes the total transmit power subject to the QoS constraint and the secrecy constraint. The QoS performance of the NOMA transmission with wiretap encoding is characterized by the confidential information rate for each user $k$, i.e., $R_{s,k}$. It is worth mentioning that in~\cite{Zhang_16_ssrNOMA} the QoS performance is captured by the codeword transmission rate for each user $k$, i.e., $R_{t,k}$, which is different from our work. We highlight that the adopted confidential information rate in this work is more suitable than the codeword transmission rate to capture the QoS performance for the transmission with the wiretap code.
This is because that the actual rate of the informative data is the confidential information rate, while the codeword transmission rate is the total rate of the informative data and the redundancy to provide secrecy. For example, a user who is served by a high codeword transmission rate with a low confidential information rate actually obtains the useful information slowly, which cannot be considered as a good quality of service.

\subsection{Problem Formulation}
The problem is formulated as follows:
\begin{subequations}\label{prob:MinSumP}
  \begin{align}\label{}
 \min_{\bm{\pi}, \bm{P}, \bm{R}_t, \bm{R}_s} &~~~  \sum_{k=1}^KP_{\pi(k)},\\
 \mathrm{s.t.}
               &~~~~ 0<P_{\pi(k)},  ~~\forall k\in\left\{1,\cdots,K\right\},\label{eq:P1_cons_ag0}\\
               &~~~~ R_{s,{\pi(k)}}\le R_{t,{\pi(k)}}, ~~\forall k\in\left\{1,\cdots,K\right\}, \label{eq:P1_cons_RsleRt}\\
               &~~~~ R_{t,{\pi(k)}}\le C_{\pi(k)}, ~~\forall k\in\left\{1,\cdots,K\right\}, \label{eq:P1_cons_tleC}\\
               &~~~~ R_{t,{\pi(k)}}\le C_{{\pi(m)},{\pi(k)}}, \forall k\in\left\{1,\cdots,K-1\right\},\notag\\ &~~~~
               \forall m\in\left\{k+1,\cdots,K\right\}, \label{eq:P1_cons_tleNOMAC}\\
               &~~~~ Q\le R_{s,{\pi(k)}}, ~~\forall k\in\left\{1,\cdots,K\right\},\label{eq:P1_QoScons}\\
               &~~~~ \psopik\le\epsilon, ~~\forall k\in\left\{1,\cdots,K\right\} \label{eq:P1_cons_pso}.
\end{align}
\end{subequations}
where $\bm{\pi}=\left[\pi(1),\cdots,\pi(K)\right]$ denotes the decoding order vector, $\bm{P}=\left[P_{\pi(k)},\cdots,P_\pi(K)\right]$ denotes the power allocation vector, $\bm{R}_t=\left[R_{t,\pi(1)},\cdots,R_{t,\pi(K)}\right]$ denotes the codeword transmission rate vector, $\bm{R}_s=\left[R_{s,\pi(1)},\cdots,R_{s,\pi(K)}\right]$ denotes the confidential information rate vector,
$C_{\pi(k)}=\log_2\left(1+\SINR_{{\pi(k)}}\right)$ denotes the channel capacity of user ${\pi(k)}$ to decode its own message, $C_{{\pi(m)},{\pi(k)}}=\log_2\left(1+\SINR_{{\pi(m)},{\pi(k)}}\right)$ denotes the channel capacity of user ${\pi(m)}$ to decode the message $s_{\pi(k)}$, $Q$ denotes the minimum acceptable confidential information rate for each user, and $0<\epsilon<1$ denotes the maximum tolerable secrecy outage probability for each message.
In the formulated problem~\eqref{prob:MinSumP}, the constraint \eqref{eq:P1_cons_tleC} is to ensure that $s_{\pi(k)}$ can be decoded by user ${\pi(k)}$ without error, and the constraint \eqref{eq:P1_cons_tleNOMAC} is to ensure that $s_{\pi(k)}$ can be decoded by user ${\pi(m)}$, $m>k$, for the SIC.
In addition, the QoS constraint is represented by~\eqref{eq:P1_QoScons}, which requires a minimum acceptable confidential information rate $Q>0$ for each user. The secrecy constraint is represented by~\eqref{eq:P1_cons_pso}, which requires a maximum tolerable secrecy outage probability $\epsilon$ for each message.
We assume that all users have the same QoS requirement $Q$ and the same secrecy
outage requirement $\epsilon$. It is worth mentioning that our analysis can be extended to the general scenario where each user have a different QoS requirement $Q_k$ and a different secrecy outage requirement $\epsilon_k$, which however is beyond the scope of this work.
Note that the problem \eqref{prob:MinSumP} is not always feasible depending on the channel conditions of the users and the eavesdropper. In this section, we assume that the problem \eqref{prob:MinSumP}  is feasible. The condition for the problem \eqref{prob:MinSumP} to be feasible will be given later in Corollary \ref{Coro:1}.

\subsection{Problem Simplification}
One can find that the optimal $R_{t,{\pi(k)}}$ is the maximum $R_{t,{\pi(k)}}$ that satisfies \eqref{eq:P1_cons_tleC} and \eqref{eq:P1_cons_tleNOMAC}.
Then, based on \eqref{eq:SINRkk} and \eqref{eq:SINRkj},  the optimal $R_{t,{\pi(k)}}$ to  the problem \eqref{prob:MinSumP} for any given decoding order $\bm{\pi}$ and power allocation $\bm{P}$ is given by
\begin{equation}\label{eq:optimalRtpik}
  R_{t,\pi(k)}^o\!=\!\left\{\!\!\begin{array}{ll}
  \log_2\left(1\!+\!\frac{\gamma_{t,\pi(k)}P_{\pi(k)}}{1+\gamma_{t,\pi(k)}\sum_{i=k+1}^{K}P_{\pi(i)}}\right)\;, &\mbox{if}~k\le K\!-\!1 \\
  \log_2\left(1\!+\!\gamma_{\pi(K)}P_{\pi(K)}\right)\;,
  &\mbox{if}~k=K,
  \end{array}
  \right.
\end{equation}
where
$\gamma_{t,\pi(k)}=\min_{i\in\left\{k,\cdots,K\right\}}\gamma_{\pi(i)}$.
One can also find that the optimal $R_{s,{\pi(k)}}$ is the minimum $R_{s,{\pi(k)}}$ that satisfies~\eqref{eq:P1_QoScons}, since the minimum power consumption is a non-decreasing function of the confidential information rate.
Then, the optimal solution of $\bm{R}_s$ to the problem \eqref{prob:MinSumP} is given by
\begin{equation}\label{eq:optimalRspik}
   R_{s,{\pi(k)}}^o=Q, ~~\forall k\in\left\{1,\cdots,K\right\}.
\end{equation}
With \eqref{eq:optimalRtpik} and \eqref{eq:optimalRspik}, the secrecy outage probability is derived as~\eqref{eq:pso_afterlemma1},  shown at the top of the next
page, where $\bar{\gamma}_e=\mathbb{E}\left\{\gamma_e\right\}=d_e^{-\alpha}/\sigma_e^2$.
\begin{figure*}[!t]
\begin{align}\label{eq:pso_afterlemma1}
  \psopik& = \mathbb{P}\left(R_{t,\pi(k)}^o-Q<\log_2\left(1+\SINR_{\tilde{\pi}(k)}\right)\right) \notag\\
  &=\left\{\begin{array}{ll}
    \exp\left(-\frac{1}{\bar{\gamma}_e}\left(\frac{\frac{1+\gamma_{t,\pi(k)}\sum_{i=k}^{K}P_{\pi(i)}}{1+\gamma_{t,\pi(k)}\sum_{i=k+1}^{K}P_{\pi(i)}}-2^{Q}}{2^{Q}\sum_{i=k}^{K}P_{\pi(i)}-\frac{1+\gamma_{t,\pi(k)}\sum_{i=k}^{K}P_{\pi(i)}}{1+\gamma_{t,\pi(k)}\sum_{i=k+1}^{K}P_{\pi(i)}}\sum_{i=k+1}^{K}P_{\pi(i)}}\right)\right)\;, &\mbox{if}~k\le K-1 \\
    \exp\left(-\frac{1}{\bar{\gamma}_e}\left(\frac{1+ \gamma_{\pi(K)}P_{\pi(K)}-2^{Q}}{2^Q P_{\pi(K)}}\right)\right)\;, &\mbox{if}~k=K.
      \end{array}
      \right.
\end{align}
\hrulefill
\end{figure*}

Based on \eqref{eq:pso_afterlemma1}, we have the following proposition to summarize the optimal decoding order to the problem \eqref{prob:MinSumP}.
\begin{Proposition}\label{prop:optdecodingpi}
When the problem~\eqref{prob:MinSumP} is feasible, the optimal decoding order $\bm{\pi}$ to the problem \eqref{prob:MinSumP} satisfies
\begin{equation}\label{eq:optimaldecodingordercondition}
  \gamma_{\pi^(1)}\le\gamma_{\pi^(2)}\cdots\le \gamma_{\pi^(K)},
\end{equation}
and hence, the optimal decoding order is given by $\bm{\pi}^o=\left[\pi^o(1),\cdots,\pi^o(K)\right]$, where
\begin{equation}\label{eq:optimaldecodingorder}
  {\pi^o(k)}=k,~~\forall k\in\left\{1,\cdots,K\right\}.
\end{equation}

\end{Proposition}
\begin{IEEEproof}
See Appendix~\ref{app:proofoptimaldecodingorder}.
\end{IEEEproof}
From Proposition~\ref{prop:optdecodingpi},
we note that the optimal decoding order for the studied NOMA scheme with the secrecy outage constraint is the same as  that for the conventional NOMA scheme without the secrecy consideration, which is in the descending order of channel gains normalized by noise.

With \eqref{eq:optimalRtpik}, \eqref{eq:optimalRspik}, \eqref{eq:optimaldecodingorder}, and the fact that \eqref{eq:P1_cons_RsleRt} always could be satisfied if \eqref{eq:P1_cons_pso} can be satisfied, the problem~\eqref{prob:MinSumP} is simplified and rewritten as:
\begin{subequations}\label{prob:MinSumP_Simple}
  \begin{align}\label{}
 \min_{\bm{ P}} &~  \sum_{k=1}^KP_k,\\
 \mathrm{s.t.} &~ 0<P_{k},  ~~\forall k\in\left\{1,\cdots,K\right\} \label{eq:P2_cons_Pk0}\\
               &~ \bar{\gamma}_e\ln\left({\epsilon^{-1}}\right)\le
               \frac{\frac{1+\gamma_k\sum_{i=k}^{K} P_i}{1+\gamma_k\sum_{i=k+1}^{K} P_i}-2^{Q}}
               {2^{Q}\sum_{i=k}^{K} P_i\!-\!\frac{1+\gamma_k\sum_{i=k}^{K} P_i}{1+\gamma_k\sum_{i=k+1}^{K} P_i}\sum_{i=k+1}^{K} P_i},\notag\\
               &~\forall k\in\left\{1,\cdots,K-1\right\}, \label{eq:P2_cons_psok}\\
                &~ \bar{\gamma}_e\ln\left({\epsilon^{-1}}\right)\le\frac{1+ \gamma_KP_K-2^{Q}}{2^{Q} P_K}. \label{eq:P2_cons_psoK}
\end{align}
\end{subequations}
Note that the problem~\eqref{prob:MinSumP_Simple} is non-convex due to the non-convex constraint in~\eqref{eq:P2_cons_psok}.
The design parameters $P_1, \cdots, P_K$ are coupled with each other in a complicated way.
Thus, the problem cannot be directly solved  by a standard convex optimization  solver~\cite{Boyd_04_ConvexOptimization}.
Furthermore, we note that the design problems in the existing studies of NOMA schemes can often be transformed into a sequence of liner programs (LPs); see, e.g., ~\cite{Timotheou_15_FairnessNOMA5G,Zhang_16_ssrNOMA,Cui_16_ANPASUOCIS}.
In contrast, our problem cannot be transformed into a sequence of LPs due to the complicated expression for the secrecy outage constraint, i.e., \eqref{eq:P2_cons_psok}. This incurs more difficulties to solve the problem \eqref{prob:MinSumP_Simple} compared with many design problems of NOMA schemes without the secrecy outage constraint.

\subsection{Optimal Solution}

The optimal solution to the problem~\eqref{prob:MinSumP_Simple} is summarized in the following theorem, based on which we can first obtain the optimal power allocated to user $K$, and then iteratively obtain the optimal power allocated to the other users.

\begin{Theorem}\label{theo:P3optalpha}
When the problem~\eqref{prob:MinSumP_Simple} is feasible, the optimal solution of $\bm{P}$ to  the problem \eqref{prob:MinSumP_Simple} is $\bm{ P}^o=\left[ P^o_1,\cdots, P^o_K\right]$, where
\begin{equation}\label{eq:P3optalpha}
      P_k^o\!\!=\!\!\left\{\!\!\!\!\!\begin{array}{ll}
     \frac{\left(\!2^{Q}\!-\!1\!\right)\left(\!1+\bar{\gamma}_e\ln\left(\!{\epsilon^{\!-\!1}}\!\right)\sum_{i=k+1}^K\!\! P^o_i\right)\left(1+\gamma_k\sum_{i=k+1}^K\!\! P^o_i\right)}
     {\gamma_k\left(\!1\!-\bar{\gamma}_e\ln\left(\!{\epsilon^{-1}}\!\right)\left(2^{Q}\!-\!1\right)\!\sum_{i=k+1}^K\! P^o_i\right)-\bar{\gamma}_e\ln\left(\!{\epsilon^{-\!1}}\!\right)2^{Q}},\!\!\!\! &\mbox{if}~\!k\!\le\! K\!\!-\!\!1 \\
  \frac{2^{Q}-1}{\gamma_K-\bar{\gamma}_e\ln\left({\epsilon^{-1}}\right)2^{Q}}, &\mbox{if}~\!k=K.
      \end{array}\right.
\end{equation}
\end{Theorem}
\begin{IEEEproof}
See Appendix~\ref{app:proofTheo1}.
\end{IEEEproof}

\begin{Corollary}\label{Coro:1}
The problem~\eqref{prob:MinSumP_Simple} is feasible if and only if
\begin{align}\label{eq:coro_kg0}
 & {\gamma_k\left(\!1\!-\!\bar{\gamma}_e\ln\left({\epsilon^{-1}}\right)\left(2^{Q}\!-\!1\right)\!\!\sum_{i=k+1}^K P^o_i\right)\!-\!\bar{\gamma}_e\ln\left({\epsilon^{-1}}\right)2^{Q}}>0,\notag\\
 &~ \forall k\in\left\{1,\cdots,K-1\right\}
\end{align}
\begin{equation}\label{eq:coro_Kg0}
  \gamma_K-\bar{\gamma}_e\ln\left({\epsilon^{-1}}\right)2^{Q}>0,
\end{equation}
where $ P_k^o$ is given in~\eqref{eq:P3optalpha}.
\end{Corollary}
\begin{IEEEproof}
See Appendix~\ref{app:proofCoro1}.
\end{IEEEproof}

\begin{Remark} According to Corollary~\ref{Coro:1}, we note that the problem~\eqref{prob:MinSumP_Simple} is not always feasible. The problem~\eqref{prob:MinSumP_Simple} is infeasible when the QoS constraint and/or the secrecy constraint cannot be satisfied even with an infinite transmit power, i.e., $\sum_{k=1}^K P_k\rightarrow\infty$. We note that the feasibility of the problem is determined by the channel condition, $\gamma_k$ and $\bar{\gamma}_e$, the QoS constraint, $Q$, and the secrecy constraint, $\epsilon$.
In practice, when it is infeasible to serve all users under the QoS and secrecy constraints, the transmitter may reduce the number of users to be served using a certain user-selection policy.
In the following, we briefly discuss some possible user-selection policies, while the detail of the user selection policy is beyond the scope of this work.

From \eqref{eq:coro_kg0} and \eqref{eq:coro_Kg0}, we find that it is infeasible to serve any user~$i$ whose channel condition  is not sufficiently good to ensure $\gamma_i>\bar{\gamma}_e\ln\left({\epsilon^{-1}}\right)2^{Q}$. Hence, the transmitter first excludes any user~$i$  with $\gamma_i\le\bar{\gamma}_e\ln\left({\epsilon^{-1}}\right)2^{Q}$. Then, the transmitter selects users among all other users. Note that the user-selection policy depends on application-specific requirements.
For example, to enhance the sum of confidential information rates, the transmitter may select the user who has the best channel condition, and iteratively adds the user whose channel condition is the best among all remaining users until \eqref{eq:coro_kg0} cannot be satisfied.
On the other hand, under fairness consideration, the transmitter may add users based on, e.g., the proportional fairness criteria until \eqref{eq:coro_kg0} cannot be satisfied.
When there is no user $k$ in the network with $\gamma_k>\bar{\gamma}_e\ln\left({\epsilon^{-1}}\right)2^{Q}$ at the time slot, the transmitter would suspend the transmission.
\end{Remark}

\section{Fairness Information Rate Maximization}\label{sec:maxFairnessRate}
In this section, we study the design problem of the NOMA scheme that maximizes the minimum confidential information rate among users subject to the secrecy constraint and an instantaneous transmit  power constraint. The objective of maximizing the minimum confidential information rate among users is motivated by the consideration of fairness among users.
We note that both design problems in this and the previous sections have taken  the secrecy constraint into account. The transmit power is considered in the objective function and the confidential information rate (the QoS performance) is considered in the constraints in Section~\ref{sec:mimpsum}, while the transmit power is considered in the constraints and the confidential information rate is considered in the objective function in this section.
In fact, we will show later that the solution to the design problem in this section  is based on the design solution given in Section~\ref{sec:mimpsum}.

\subsection{Problem Formulation}
The problem is formulated as follows:
\begin{subequations}\label{prob:MaxminRs}
  \begin{align}\label{}
 \max_{\bm{\pi}, \bm{P}, \bm{R}_t, \bm{R}_s} &~~~  \min_{k\in\left\{1,\cdots,K\right\}}R_{s,\pi(k)},\\
 \mathrm{s.t.} &~~~~ \eqref{eq:P1_cons_ag0}, \eqref{eq:P1_cons_RsleRt}, \eqref{eq:P1_cons_tleC}, \eqref{eq:P1_cons_tleNOMAC}, \eqref{eq:P1_cons_pso},\\
  &~~~~ \sum_{k=1}^KP_{\pi(k)}\le P,\label{eq:P1_cons_suma}
 \end{align}
\end{subequations}
where \eqref{eq:P1_cons_pso} and~\eqref{eq:P1_cons_suma} represent the secrecy constraint and the power constraint, respectively, and $P>0$ denotes the maximum available instantaneous transmit power. 

Following the similar analysis in Section~\ref{sec:mimpsum}, the optimal solution of $\bm{\pi}$, $\bm{R}_t$, and $\bm{R}_s$, to  the problem \eqref{prob:MaxminRs} are given by
\begin{equation}\label{eq:}
  {\pi^*(k)}=k,~~\forall k\in\left\{1,\cdots,K\right\},
\end{equation}
\begin{equation}\label{eq:}
  R_{t,k}^*=\left\{\begin{array}{ll}
  \log_2\left(1+\frac{\gamma_{k}P_{k}}{1+\gamma_{k}\sum_{i=k+1}^{K}P_{i}}\right)\;, &\mbox{if}~k\le K-1 \\
  \log_2\left(1+\gamma_{K}P_{K}\right)\;,
  &\mbox{if}~k=K,
  \end{array}
  \right.
\end{equation}
\begin{equation}\label{eq:P2optimalRsk}
   R_{s,k}^*=\Rsmin, ~~\forall k\in\left\{1,\cdots,K\right\},
\end{equation}
respectively.
The problem~\eqref{prob:MaxminRs} is then simplified as:
\begin{subequations}\label{prob:MaxminRs_simple}
  \begin{align}\label{}
  \max_{\bm{ P}, R_{s,\min}} &~~~  R_{s,\min},\\
 \mathrm{s.t.} &~~~~ \eqref{eq:P2_cons_Pk0}~\text{and}~\eqref{eq:P1_cons_suma},\\
               &~~~~ 0<\Rsmin,\label{eq:P2_Rsming0}\\
               &~~~~ \eqref{eq:P2_cons_psok} ~\text{with}~ Q=\Rsmin, \label{eq:Pmaxmin_cons_psok}\\
                &~~~~ \eqref{eq:P2_cons_psoK} ~\text{with}~ Q=\Rsmin. \label{eq:Pmaxmin_cons_psoK}
\end{align}
\end{subequations}
Note that the constraint \eqref{eq:Pmaxmin_cons_psok} is still non-convex. The design parameters $P_1, \cdots, P_K$ and $\Rsmin$ are coupled with each other in a complicated way. 
\subsection{Optimal Solution}

\begin{Proposition}\label{prop:feasiblePmaxmin}
 The problem \eqref{prob:MaxminRs_simple} is feasible if and only if
 \begin{equation}\label{eq:feasibleconditionproblemMaxminRs_simple}
   \gamma_k>\bar{\gamma}_e\ln\left(\epsilon^{-1}\right), ~~\forall k\in\left\{1,\cdots,K\right\}.
 \end{equation}
\end{Proposition}
\begin{IEEEproof}
See Appendix~\ref{app:prooffeasiblePmaxmin}.
\end{IEEEproof}

\begin{Remark}
According to Proposition~\ref{prop:feasiblePmaxmin}, the problem \eqref{prob:MaxminRs_simple} is not always feasible. The problem~\eqref{prob:MaxminRs_simple} is infeasible when a positive minimum confidential information rate, i.e., $\Rsmin>0$, is never achievable subject to the secrecy constraint.
The feasibility of the problem \eqref{prob:MaxminRs_simple} is determined by the channel condition, $\gamma_k$ and $\bar{\gamma}_e$, and the secrecy constraint, $\epsilon$.
To ensure that the transmission design is feasible, the transmitter can adopt a simple user-selection policy to exclude any user~$i$ whose channel condition is not sufficiently good such that $\gamma_i\le\bar{\gamma}_e\ln\left({\epsilon^{-1}}\right)$.
When there is no user~$k$ in the network with $\gamma_k>\bar{\gamma}_e\ln\left({\epsilon^{-1}}\right)$ at the time slot, the transmitter would suspend the transmission.
In the rest of the analysis in this section, we assume that \eqref{eq:feasibleconditionproblemMaxminRs_simple} is satisfied.
\end{Remark}

In the following, we first describe the basic methodology to solve the problem \eqref{prob:MaxminRs_simple}. Then, we present an iterative algorithm to solve the problem. We further derive the closed-form solution to the problem for the case of two users.

The methodology to solve the problem \eqref{prob:MaxminRs_simple} is given as follows.
Denote the optimal $\Rsmin$ to the problem~\eqref{prob:MaxminRs_simple} as $\Rsmin^*$. We can easily find that $\Rsmin^*$ is unique. For a constant value $Q$,
if the following problem~\eqref{prob:interProbchangemaxminto} is feasible, then $\Rsmin^*\ge Q$. Otherwise, $\Rsmin^*<Q$.
\begin{subequations}\label{prob:interProbchangemaxminto}
  \begin{align}\label{}
 \min_{\bm{ P}} &~~~  \sum_{k=1}^KP_k, \\
 \mathrm{s.t.} &~~~~ \eqref{eq:P2_cons_Pk0}, \eqref{eq:P2_cons_psok}, \eqref{eq:P2_cons_psoK},~\text{and}~\eqref{eq:P1_cons_suma}.
\end{align}
\end{subequations}
The feasibility problem of~\eqref{prob:interProbchangemaxminto} is hard to solve directly. Instead, we transform the feasibility problem of~\eqref{prob:interProbchangemaxminto} to the problem \eqref{prob:MinSumP_Simple} in Section~\ref{sec:mimpsum}. The problem~\eqref{prob:interProbchangemaxminto} is feasible if and only if the problem \eqref{prob:MinSumP_Simple} is feasible and the solution to the problem \eqref{prob:MinSumP_Simple} satisfies~\eqref{eq:P1_cons_suma}, i.e., $\sum_{k=1}^KP^o_k\le P$, where $P^o_k$ is given in \eqref{eq:P3optalpha}.
From the analysis in Section~\ref{sec:mimpsum}, both the feasibility problem of  \eqref{prob:MinSumP_Simple} and the problem~\eqref{prob:MinSumP_Simple} itself are solvable.
Thus, the feasibility problem of \eqref{prob:interProbchangemaxminto} is solvable.

For an arbitrary number of users, we can obtain the optimal $\Rsmin$ and $\bm{P}$ with a bisection search as summarized in Algorithm~\ref{Alg:1}.
We highlight that the complexity of Algorithm~\ref{Alg:1} is low. Because the error of a bisection search algorithm decreases exponentially as the number of iterations increases, the required number of iterations to achieve an acceptable error $v$ is relatively small. In addition, the complexity of each iteration in Algorithm~\ref{Alg:1} is low, since the associated inner problem (feasibility problem) of each iteration has the closed-form solution, i.e., \eqref{eq:P3optalpha}. The detailed analysis on the complexity of Algorithm~\ref{Alg:1} is given as follows.
For a given acceptable error $v$, the number of iterations of Algorithm~\ref{Alg:1}  is equal to $\left\lceil\log_2\left(\frac{Q^{\UB}- Q^{\LB}}{v}\right)\right\rceil$.
We note that the complexity of each iteration is dominated by the derivations of \eqref{eq:P3optalpha}. Based on the rules of counting floating point operations (flops)~\cite{Thant_05_FlopsCount}, we obtain the complexity of \eqref{eq:P3optalpha} as $O\left(1.5K^2+24.5K-13\right)$ flops. Then, the total complexity of the algorithm is equal to the product of
the number of iterations and the complexity of each iteration, which is given by $O\left(\left(1.5K^2+24.5K-13\right)\left\lceil\log_2\left(\frac{Q^{\UB}- Q^{\LB}}{v}\right)\right\rceil\right)$.

\begin{algorithm}[!htb]
\caption{Optimal solution to the problem~\eqref{prob:MaxminRs_simple}}
\begin{algorithmic}[1]\label{Alg:1}
\STATE{\textbf{input}
~~Channel condition: $\gamma_1,\cdots,\gamma_K$ and $\gamma_e$;\\
~\quad\quad\quad Transmit power constraint: $P$;\\
~\quad\quad\quad Secrecy constraint: $\epsilon$;\\
~\quad\quad\quad Acceptable error of $\Rsmin^*$: $v$ (e.g., $v=10^{-10}$);\\
\STATE{\textbf{output}
~\!Optimal rate: $ \Rsmin^*$;\\
~~~~~~~~~~\!\!\! Optimal power allocation parameters: $\bm{ P}^*=\left[ P^*_1,\cdots, P^*_K\right]$;
}}
\STATE{Initialize $ Q^{\LB}=0$; $ Q^{\UB}=\log_2\left(1+\gamma_1P\right)$;}
\WHILE{$ Q^{\UB}- Q^{\LB}\ge v$}
\STATE{Set $ Q=\left( Q^{\UB}+ Q^{\LB}\right)/2$;}
\STATE{Obtain $\bm{ P}^o=\left[ P^o_1,\cdots, P^o_K\right]$ according to \eqref{eq:P3optalpha};}
\IF{\eqref{eq:coro_kg0}, \eqref{eq:coro_Kg0}, and $\sum_{k=1}^KP^o_k\le P$ are all satisfied} 
\STATE{Set $ Q^{\LB}= Q$;} \COMMENT{Update the upper bound of $\Rsmin^*$.}
\STATE{Set $ \Rsmin^*= Q$;}
\STATE{Set $\bm{ P}^*=\bm{ P}^o$;}
\ELSE
\STATE{Set $ Q^{\UB}= Q$;} \COMMENT{Update the upper bound of $\Rsmin^*$.}
\ENDIF
\ENDWHILE
\RETURN{$ \Rsmin^*$ and $\bm{ P}^*$;}
\COMMENT{The solution to the problem \eqref{prob:MaxminRs_simple} is obtained.}
\end{algorithmic}
\end{algorithm}

In the case of two users, we obtain the closed-form expressions for the solutions to the problem~\eqref{prob:MaxminRs_simple}, which are summarized in the following Proposition~\ref{prop:1twousersolutio}. It is worth mentioning that the scenario of two users
to perform NOMA jointly is of practical interest. Since a NOMA system is strongly interference-limited, having a large number of users to perform NOMA jointly is usually not realistic.
Especially, it is often to group two users together to perform NOMA jointly in the studies of NOMA schemes with the technique of user pairing, e.g., \cite{Ding_16_IuserpairingNMADT}.

\begin{Proposition}\label{prop:1twousersolutio}
When the problem \eqref{prob:MaxminRs_simple} is feasible, the optimal solutions of $\bm{P}$ and $ \Rsmin$ to the problem~\eqref{prob:MaxminRs_simple} with $K=2$ are given by
\begin{align}\label{eq:a1starprop1}
  P_1^*=& \frac{\left(1+\phi P\right)\left(\gamma_2+\gamma_1\left(1+2\gamma_2P\right)-2\phi\left(1+\gamma_1P\right)\right)-\psi}
 {2\left(\left(1+\phi P\right)\gamma_1\gamma_2-\phi^2\left(1+\gamma_1 P\right)\right)},
\end{align}
\begin{equation}\label{eq:a2starprop1}
   P_2^*=\frac{\psi-\left(\gamma_1+\gamma_2\right)-\phi\left(\gamma_2P-\gamma_1P-2\right)}
  {2\left(\left(1+\phi P\right)\gamma_1\gamma_2-\phi^2\left(1+\gamma_1 P\right)\right)},
\end{equation}
and
 \begin{equation}\label{eq:Rsstarprop1}
    \Rsmin^*=\frac{\psi-\left(1+\phi P\right)\left(\gamma_2-\gamma_1\right)}{2\left(1+\phi P\right)\left(\gamma_1-\phi\right)},
 \end{equation}
 respectively,
where $\phi=\bar{\gamma}_e\ln\left({\epsilon^{-1}}\right)$ and \\
 \begin{small} $\psi\!=\!\!\sqrt{\!\left(1\!+\!\phi P\right)\left(4\left(1\!+\!\gamma_1 P\right)\left(\gamma_1\!-\!\phi\right)\left(\gamma_2\!-\!\phi\right)\!+\!\left(1\!+\!\phi P\right)\left(\gamma_2\!-\!\gamma_1\right)^2\right)}$.\end{small}
\end{Proposition}
\begin{IEEEproof}
See Appendix~\ref{app:proofprop1}.
\end{IEEEproof}

Denote the optimal ratio of power allocated to user 1 as
\begin{equation}\label{eq:p1optdividedbyP}
  \beta_1^*=\frac{P_1^*}{P}=\frac{\left(1\!+\!\phi P\right)\left(\gamma_2\!+\!\gamma_1\left(1\!+\!2\gamma_2P\right)\!-\!2\phi\left(1\!+\!\gamma_1P\right)\right)\!-\!\psi}
 {2P\left(\left(1+\phi P\right)\gamma_1\gamma_2-\phi^2\left(1+\gamma_1 P\right)\right)}.
\end{equation}
From Proposition~\ref{prop:1twousersolutio}, we have the following Corollary.
\begin{Corollary}\label{Coro:beta1increase}
The optimal ratio of power allocated to user 1, $\beta_1^*$, is a monotonously increasing function of $\phi=\bar{\gamma}_e\ln\left({\epsilon^{-1}}\right)$. 
\end{Corollary}
\begin{IEEEproof}
See Appendix~\ref{app:proofbeta1increase}.
\end{IEEEproof}

From Corollary~\ref{Coro:beta1increase}, we find that
$  {\partial \beta_1^*}/{\partial\bar{\gamma}_e}>0$
and
$ {\partial \beta_1^*}/{\partial\epsilon}<0.$
Thus,  it is wise to increase the ratio of power allocated to the user whose channel condition is bad, i.e., user 1, when the eavesdropper's channel condition improves and/or the secrecy constraint becomes more stringent.

\section{Analytical Comparison Between NOMA and OMA}\label{sec:analyticalcomp}
In this section, we analytically compare the performance of the NOMA scheme and a benchmark OMA scheme.
We focus on the performance comparison in terms of the maximum achievable minimum confidential information rate among users subject to the secrecy outage constraint and the transmit power constraint, i.e., the problem \eqref{prob:MaxminRs} in Section~\ref{sec:maxFairnessRate}.
The performance comparison in terms of the minimum transmit power required to satisfy the QoS constraint and the secrecy constraint is similar, and hence, the details are omitted for the sake of brevity.

%
%


\subsection{Benchmark Scheme}
We consider the TDMA scheme as a benchmark, which is a typical OMA scheme.
Note that we can also consider the orthogonal frequency-division multiple access (OFDMA) as the benchmark OMA scheme. The bandwidth resources are split between users for the OFDMA and the time resources are split between users for the TDMA. In fact, the analyses of the OFDMA and the TDMA as the benchmark are mathematically equivalent in this paper~\cite{Ding_16_appNMALTE5G,Saito_13_N-OMAforCFRA}.
As per the mechanism of the TDMA scheme, the transmission during each fading block is divided into $K$ sub-slots. The $k$-th sub-slot is used to serve user $k$.
With the fixed transmit power $P$, the received SNRs at user $k$ and the eavesdropper in the $k$-th sub-slot are given by $\gamma_kP$ and $\gamma_eP$, respectively.
Denote the time ratio allocated to user $k$ as $t_k$, $\sum_{k=1}^Kt_k\le1$.
The effective channel capacity for user $k$ and the eavesdropper to decode the message $s_k$ are given by
\begin{equation}\label{}
  C_k^{\TDMA}=t_k\log_2\left(1+\gamma_kP\right)
\end{equation}
and
\begin{equation}\label{}
  C_{e,k}^{\TDMA}=t_k\log_2\left(1+\gamma_eP\right),
\end{equation}
respectively.
It is easy to find that the optimal codeword transmission rate for the message $s_k$ is equal to the user capacity, i.e.,
\begin{equation}\label{}
  R_{t,k}^{\TDMA}=C_k^{\TDMA}=t_k\log_2\left(1+\gamma_kP\right).
\end{equation}
Subject to the secrecy outage constraint
$\psok\le\epsilon$, the maximum confidential information rate for the message $s_k$ is then limited by
$
t_k\left[\log_2\left(1+P\gamma_k\right)-\log_2\left(1+P\bar{\gamma}_e\ln\left(\epsilon^{-1}\right)\right)\right]^+,
$
where $[x]^+=\max\left\{x,0\right\}$.
Note that the condition of ensuring a positive confidential information rate for all users is that
$\gamma_k>\bar{\gamma}_e\ln\left(\epsilon^{-1}\right)$, $\forall k\in\left\{1,\cdots,K\right\}$, which is the same as the condition for the NOMA scheme as given in Proposition~\ref{prop:feasiblePmaxmin}. 
Thus, the achievable confidential information rate for user $k$ is given by
\begin{equation}\label{eq:OMARsk}
  R_{s,k}^{\TDMA}\le t_k\log_2\left(\frac{1+P\gamma_k}{1+P\bar{\gamma}_e\ln\left(\epsilon^{-1}\right)}\right),
\end{equation}
and the minimum confidential information rate among all users is given by
\begin{equation}\label{}
   \Rsmin^{\TDMA}=\min_{k\in\left\{1,\cdots,K\right\}}R_{s,k}^{\TDMA}.
\end{equation}

It is worth mentioning that most of the existing studies on  NOMA adopted the conventional TDMA scheme with equal time sharing, i.e., $t_1=\cdots=t_K$, as the benchmark,  e.g.,~\cite{Ding_14_CoopNonMA5S,Ding_16_appNMALTE5G,Zhang_16_ssrNOMA,Cui_16_ANPASUOCIS}.
In this section, we consider the TDMA scheme with designable time allocation ratios as the benchmark, and will consider the conventional TDMA scheme with equal time sharing later in the numerical results in Section~\ref{Sec:NumericalResults}.
In fact, the conventional TDMA scheme with equal time sharing is a special case of the TDMA scheme with designable time allocation ratios, $t_1, \cdots, t_k$. Therefore, if the NOMA scheme outperforms the TDMA scheme with designable time allocation ratios, then the NOMA scheme also outperforms the conventional TDMA scheme with equal time sharing.


\subsection{Comparison Result}
We focus on the case of two users, which is of practical interest as mentioned before. 
For the NOMA scheme, the maximum achievable minimum confidential information rate among users is $ \Rsmin^*$ given in~\eqref{eq:Rsstarprop1}.
For the TDMA scheme, we can obtain the optimal time allocation that maximizes $ \Rsmin^{\TDMA}$ by solving  for $t_1$ and $t_2$ in the equations $R_{s,1}^{\TDMA}=R_{s,2}^{\TDMA}$ and $t_1+t_2=1$, and the maximum $ \Rsmin^{\TDMA}$ is given by
\begin{equation}\label{}
  \max_{t_1, t_2} \Rsmin^{\TDMA}=\frac{\log_2\left(\frac{1+\gamma_1P}{1+\bar{\gamma}_e\ln\left(\epsilon^{-1}\right)P}\right)\log_2\left(\frac{1+\gamma_2P}{1+\bar{\gamma}_e\ln\left(\epsilon^{-1}\right)P}\right)}
  {\log_2\left(\frac{\left(1+\gamma_1P\right)\left(1+\gamma_2P\right)}{\left(1+\bar{\gamma}_e\ln\left(\epsilon^{-1}\right)P\right)^2}\right)}.
\end{equation}

The comparison result between the NOMA scheme and the TDMA scheme is summarized in the following proposition.
\begin{Proposition}\label{prop:propNOMAOMAc}
When $\gamma_1\neq\gamma_2$, the NOMA scheme can always achieve a higher maximum minimum confidential information rate among users than that of the TDMA scheme, i.e.,
\begin{equation}\label{}
     \Rsmin^*  > \max_{t_1, t_2}  \Rsmin^{\TDMA},~~ \text{if} ~~\gamma_1\neq\gamma_2.
\end{equation}
When $\gamma_1=\gamma_2$, the  NOMA scheme and the TDMA scheme can  achieve the same maximum minimum confidential information rate among users, i.e.,
\begin{equation}\label{}
     \Rsmin^*   = \max_{t_1, t_2}  \Rsmin^{\TDMA},~~ \text{if} ~~\gamma_1=\gamma_2.
\end{equation}
\end{Proposition}


\begin{IEEEproof}
See Appendix~\ref{app:proofpropNOMAOMAc}.
\end{IEEEproof}

\section{Numerical Results}\label{Sec:NumericalResults}
\begin{figure}[!htb]
\centering
\vspace{-0mm}
\includegraphics[width=.9\columnwidth]{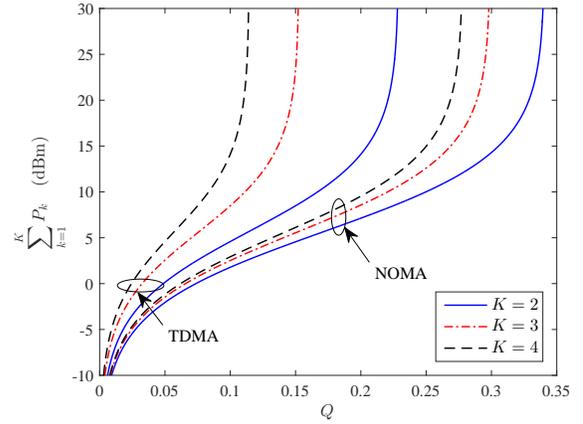}
\vspace{-0mm}
\caption{Minimum transmit power versus QoS requirement. The parameters are $\gamma_k=23+2k$~dB, $\bar{\gamma}_e=20$ dB, and $\epsilon=0.1$.}
\vspace{-0mm}  \label{fig:InstPsumVsQ}
\end{figure}

\begin{figure}[!htb]
\centering
\vspace{-0mm}
\includegraphics[width=.9\columnwidth]{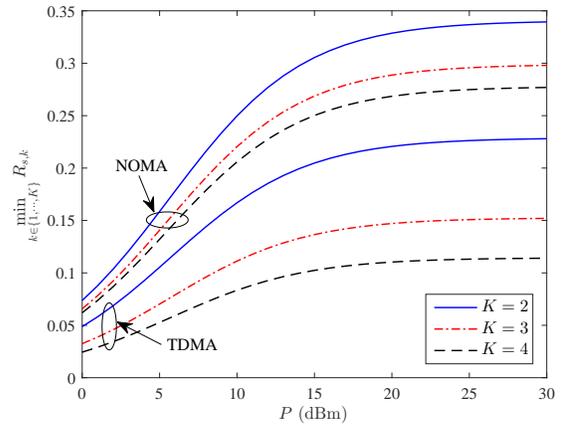}
\vspace{-0mm}
\caption{Maximum minimum confidential information rate among users versus transmit power constraint. The parameters are $\gamma_k=23+2k$~dB, $\bar{\gamma}_e=20$ dB, and $\epsilon=0.1$.}
\vspace{-4mm}  \label{fig:InstRsminVsP}
\end{figure}

\begin{figure}[!t]
\centering
\vspace{-0mm}
\includegraphics[width=.9\columnwidth]{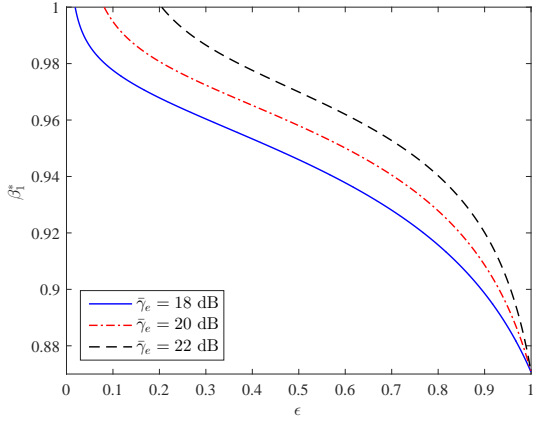}
\vspace{-0mm}
\caption{Optimal ratio of power allocated to user 1 versus secrecy outage constraint. The parameters are $P=20$ dBm, $\gamma_1=24$~dB, and $\gamma_2=26$~dB.}
\vspace{-0mm}  \label{fig:beta1vseps}
\end{figure}

\begin{figure}[!t]
\centering
\vspace{-0mm}
\includegraphics[width=.0\columnwidth]{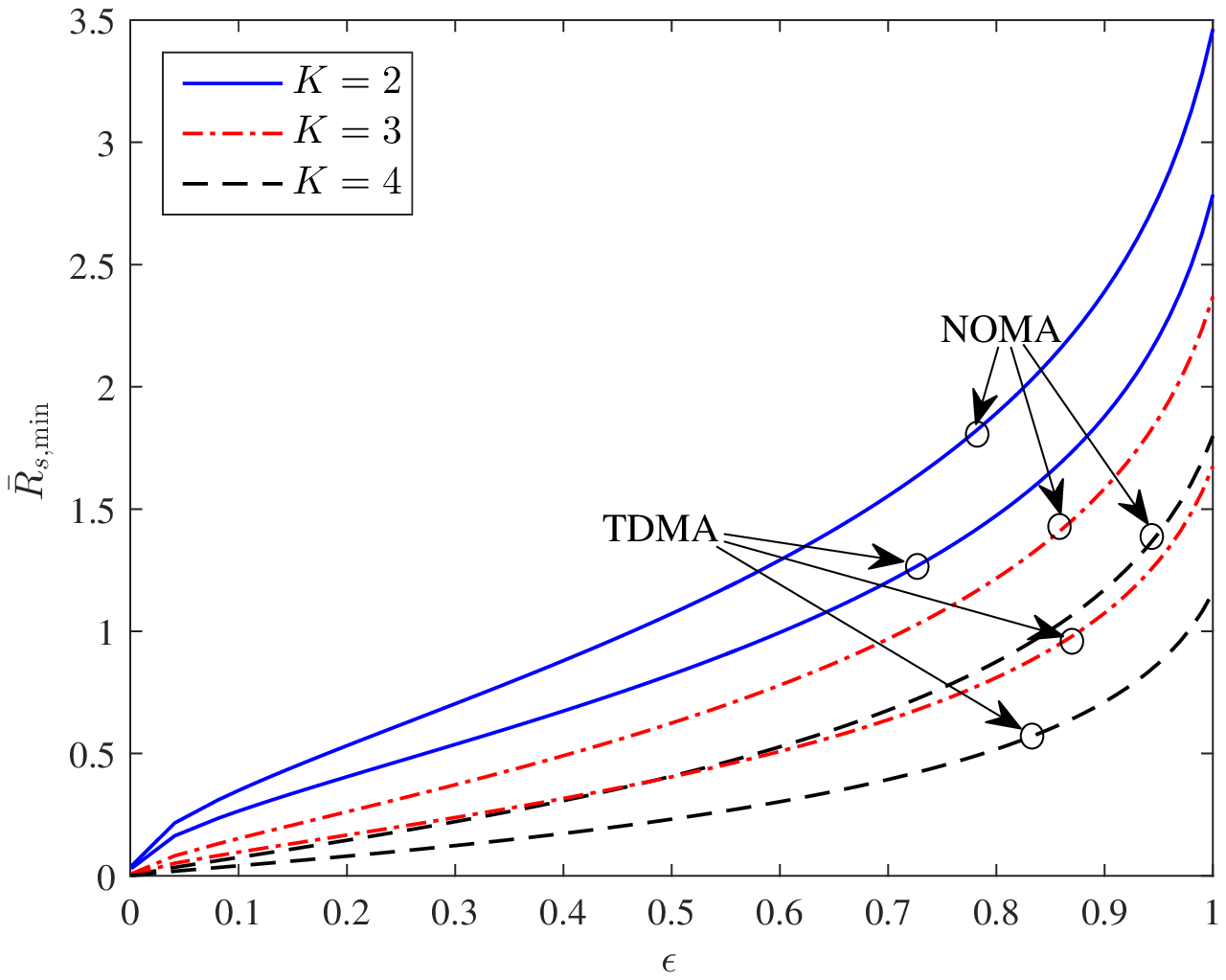}
\vspace{-0mm}
\caption{Average maximum minimum confidential information rate among users versus transmit power constraint. The parameters are $P=20$ dBm, $\alpha=4$, $d_k=50$, $d_e=80$, and $\sigma_u^2=\sigma_e^2=-70$ dBm.}
\vspace{-0mm}  \label{fig:FadingRsminVsEps}
\end{figure}

\begin{figure}[htb!]
\vspace{-0mm}
    \centering
    \begin{subfigure}[t]{.9\columnwidth}
        \centering
        \includegraphics[width=\columnwidth]{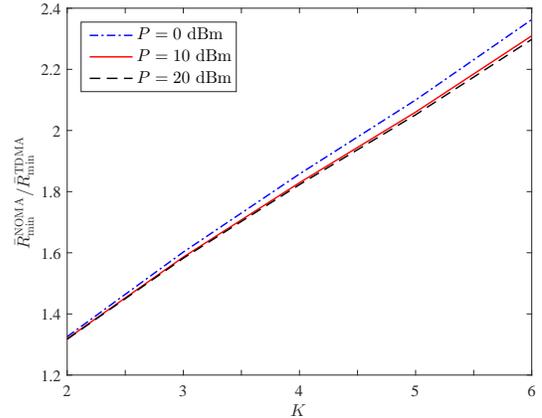}
        \caption{$d_k=50$.} \vspace{0mm}
    \end{subfigure}
    \\
      \begin{subfigure}[t]{0.9\columnwidth}
        \centering
        \includegraphics[width=\columnwidth]{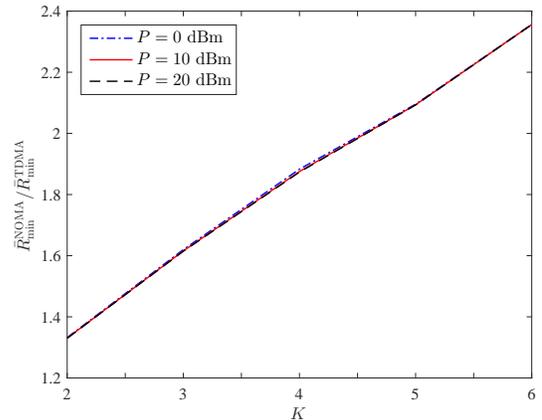}
        \caption{$d_k=80$.} \vspace{0mm}
    \end{subfigure}
    \caption{Performance gain of NOMA over  TDMA versus number of users. The parameters are $\epsilon=0.1$, $\alpha=4$, $d_e=80$, and $\sigma_u^2=\sigma_e^2=-70$ dBm.} \vspace{-0mm}\label{fig:FadingGainVsK}
\end{figure}

In this section, we present the numerical results for the designed NOMA schemes with the secrecy outage constraint.
We evaluate the NOMA scheme under a given channel realization  in Figures~\ref{fig:InstPsumVsQ}, \ref{fig:InstRsminVsP}, and \ref{fig:beta1vseps}, 
and show the average performance of the NOMA scheme over 50,000 randomly generated channel realizations in Figures~\ref{fig:FadingRsminVsEps} and \ref{fig:FadingGainVsK}. 

We first illustrate the advantage of the NOMA scheme over the OMA scheme in Figures \ref{fig:InstPsumVsQ} and~\ref{fig:InstRsminVsP}. Figure~\ref{fig:InstPsumVsQ} depicts the performance in minimizing the transmit power, $\sum_{k=1}^KP_{k}$, subject to the QoS constraint, $Q$, which is the problem studied in Section~\ref{sec:mimpsum}. Figure~\ref{fig:InstRsminVsP} shows the performance in maximizing the minimum confidential information rate among users, $\min_{k\in\left\{1,\cdots,K\right\}}R_{s,k}$, subject to the transmit power constraint, $P$, which is the problem studied in Section~\ref{sec:maxFairnessRate}. For both figures, the secrecy outage constraint is fixed at $\epsilon=0.1$. Following the existing studies on NOMA, e.g.,~\cite{Ding_14_CoopNonMA5S,Ding_16_appNMALTE5G,Zhang_16_ssrNOMA,Cui_16_ANPASUOCIS}, we adopt the conventional TDMA as the benchmark OMA scheme.
As shown in both figures, the NOMA scheme always outperforms the  TDMA scheme for different numbers of users, $K$.
As depicted in Figure~\ref{fig:InstPsumVsQ}, for both the NOMA scheme and the TDMA scheme, $\sum_{k=1}^KP_{k}$ increases as $Q$ increases, and the infeasible $Q$ cannot be satisfied even with an infinite transmit power, i.e., $\sum_{k=1}^KP_{k}\rightarrow\infty$. The feasible $Q$ for the NOMA scheme can be found according to Corollary~\ref{Coro:1}.
We note that the designed NOMA scheme achieves a better QoS for all users than that for the TDMA scheme.
From Figure~\ref{fig:InstRsminVsP}, one can note that $\min_{k\in\left\{1,\cdots,K\right\}}R_{s,k}$ increases as $P$ increases, and  $\min_{k\in\left\{1,\cdots,K\right\}}R_{s,k}$ approaches an upper bound as $P$ continues to increase. For the designed NOMA scheme, this upper bound  is actually equal to the maximum feasible $Q$ to the problem studied in Section~\ref{sec:mimpsum}.


We then demonstrate the optimal power allocation strategy subject to different secrecy outage constraints.
We consider the power allocation ratio among two users to maximize the minimum confidential information rate among users subject to a given transmit power constraint of $P=20$ dBm.
Figure~\ref{fig:beta1vseps} plots the optimal ratio of power allocated to user 1, $\beta_1^*=P_1^*/P$ versus the secrecy outage constraint, $\epsilon$. As shown in the figure, $\beta_1^*$ monotonously decreases as $\epsilon$ increases. Comparing different curves, we find that $\beta_1^*$ increases as $\bar{\gamma}_e$ increases. These observations are consistent with the analytical result given in Corollary~\ref{Coro:beta1increase}. Note that user 1 represents the user who has a relatively bad channel. Thus, it is wise to reduce the power allocated to the weak user, when the secrecy constraint becomes loose, while it is wise to increase the power allocated to the weak user, when the eavesdropper's channel condition improves.

We now present the tradeoff between the secrecy performance and the QoS performance of the NOMA scheme. Figure~\ref{fig:FadingRsminVsEps} plots the average minimum confidential information rate among users over different randomly generated channel realizations, which is denoted by $\bar{R}_{s,\min}=\mathbb{E}\left\{\min_{k}R_{s,k}\right\}$, versus the secrecy outage constraint, $\epsilon$. The transmit power constraint is fixed at $P=20$ dBm.
As shown in the figure, for both the NOMA scheme and the  TDMA scheme, there is  a clear tradeoff between the QoS performance and the secrecy performance. A better QoS performance can be obtained at a cost of secrecy performance, and vice versa, since $\bar{R}_{s,\min}$ increases as $\epsilon$ increases.  Comparing the NOMA scheme and the  TDMA scheme, we note that the NOMA scheme always achieves a better QoS-secrecy tradeoff than that of the  TDMA scheme regardless of how many users to be served.

Finally, we show the performance gain of our proposed NOMA scheme over the benchmark OMA scheme.
Figure \ref{fig:FadingGainVsK} plots $\bar{R}_{\min}^{\rm{NOMA}}/\bar{R}_{\min}^{\rm{TDMA}}$ versus the number of users, $K$, where $\bar{R}_{\min}^{\rm{NOMA}}$ and $\bar{R}_{\min}^{\rm{TDMA}}$ denote the average maximum minimum confidential information rates among users achieved by the NOMA scheme and the  TDMA scheme, respectively.
Figure~\ref{fig:FadingGainVsK}(a) demonstrates the results for the scenario where the users have better channel statistics than the eavesdropper  ($d_k<d_e$). Figure~\ref{fig:FadingGainVsK}(b) demonstrates the results for the scenario where the users have the same channel statistics as the eavesdropper ($d_k=d_e$).
As illustrated in both figures, the value of $\bar{R}_{\min}^{\rm{NOMA}}/\bar{R}_{\min}^{\rm{TDMA}}$ is always larger than 1, which indicates that our proposed NOMA scheme always has the advantage over the TDMA scheme. Comparing Figures~\ref{fig:FadingGainVsK}(a)  and~\ref{fig:FadingGainVsK}(b), we find that having the same channel statistics between the users and the eavesdropper does not affect the performance gain achieved by the NOMA scheme over the TDMA scheme.
As shown in both figures, $\bar{R}_{\min}^{\rm{NOMA}}/\bar{R}_{\min}^{\rm{TDMA}}$ increases almost linearly as $K$ increases, which indicates that the performance gain becomes more significant as the number of users increases.

\section{Conclusion and Future Work}\label{sec:conclus}
In this paper, we have studied the optimal design of secure NOMA schemes. We have first designed the NOMA scheme that minimizes the transmit power subject to the secrecy outage and QoS constraints. We have then designed the NOMA scheme that maximizes the minimum confidential information rate subject to the secrecy outage and transmit power constraints.
The optimal solutions show that the optimal decoding order for NOMA does not change by having the secrecy outage constraint in the studied problems and one should increase the ratio of power allocated to the weak user when the secrecy constraint becomes more stringent. Numerical results have  demonstrated that the NOMA scheme outperforms the OMA scheme  regardless of how many users to be served.
One interesting future research direction is to introduce the MIMO technique to secure NOMA systems.
In addition, this work has focused on the secrecy issue against the external eavesdropper in NOMA systems. The secrecy issue among users in fact is more challenging to address in NOMA systems, since most users need to decode other users' signals before decoding their own signals for the SIC. Thus, how to ensure the secrecy among users in NOMA systems is another interesting future research direction.
Furthermore, how to enhance the secrecy performance of the  code domain NOMA is an important problem to address, while our work has considered the power domain NOMA only.

\appendices
\section{Proof of Proposition~\ref{prop:optdecodingpi}}\label{app:proofoptimaldecodingorder}
Consider any given power allocation vector $\bm{P}=\left[P_{\pi(k)},\cdots,P_\pi(K)\right]$. Assume that there is an arbitrary decoding order with user $\pi(x)=m$ and user $\pi(x+1)=n$, where $\gamma_{m}>\gamma_{n}$. If we switch the decoding order of these two users such that $\pi(x)=n$ and $\pi(x+1)=n$, we find that any $\gamma_{t,\pi(k)}=\min_{i\in\left\{k,\cdots,K\right\}}\gamma_{\pi(i)}$, $k\neq x+1$, will not change, while  $\gamma_{t,\pi(x+1)}$ will increase or remain unchanged.
From~\eqref{eq:pso_afterlemma1}, we find that  $\psopik$ decreases as $\gamma_{t,\pi(k)}$ increases. Thus, switching these two users does not change the secrecy outage probability for any message $s_{\pi(k)}$, $k\neq m+1$, while may decrease the secrecy outage probability for the message $s_{\pi(m+1)}$.

The discussion above shows that, subject to any given power allocation, switching the decoding order of these two users does not worse the performance of the NOMA scheme but may improve the performance. Hence, it is wise to iteratively switch the decoding order of any two users $\pi(x)$ and $\pi(x+1)$ with $\gamma_{\pi(x)}>\gamma_{\pi(x+1)}$ until we get the optimal decoding order, which satisfies \eqref{eq:optimaldecodingordercondition}.
Note that $\gamma_1\le\cdots\le\gamma_K$. Then, $\bm{\pi}^o$ given in Proposition~\ref{prop:optdecodingpi}  is the optimal decoding order to the problem.
This completes the proof.

\section{Proof of Theorem \ref{theo:P3optalpha}}\label{app:proofTheo1}
Rewrite \eqref{eq:P2_cons_psok} as $\bar{\gamma}_e\ln\left({\epsilon^{-1}}\right)\le f_k\left(x,y\right)$, $\forall k\in\left\{1,\cdots,K-1\right\}$, where $x= P_k$, $y=\sum_{i=k+1}^K P_i$, and
\begin{equation}\label{}
  f_k\left(x,y\right)=
  \frac{\frac{1+\gamma_k (x+y)}{1+\gamma_ky}-2^{Q}}
  {2^{Q}(x+y)-\frac{1+\gamma_k (x+y)}{1+\gamma_ky}y}.
\end{equation}
Taking the partial derivative of $f_k\left(x,y\right)$ with respect to $x$, we have
\begin{equation}\label{}
  \frac{\partial f_k\left(x,y\right)}{\partial x}=\frac{\left(2^{Q}-1\right)2^{Q}\left(1+\gamma_k y\right)^2}{\left(2^{Q}\left(x\!+\!y\right)\left(1\!+\!\gamma_ky\right)\!-\!y\left(1\!+\!\gamma_k\left(x\!+\!y\right)\right)\right)^2}>0.
\end{equation}
Taking the partial derivative of $f_k\left(x,y\right)$ with respect to $y$, we have
\begin{equation}\label{}
   \frac{\partial f_k\left(x,y\right)}{\partial y}=\frac{\left(2^{Q}-1\right)\left(2^{Q}\left(1+\gamma_ky\right)^2-\left(1+\gamma_k\left(x+y\right)\right)^2\right)}
   {\left(y-\left(x+y\right)\left(s+\gamma_k\left(2^{Q}-1\right)y\right)\right)^2}.
\end{equation}
We note that $\text{sgn}\left\{\frac{\partial f_k\left(x,y\right)}{\partial y}\right\}=\text{sgn}\left\{2^{Q}\left(1+\gamma_ky\right)^2-\left(1+\gamma_k\left(x+y\right)\right)^2\right\}$. From the constraint
\begin{equation}\label{}
Q\le R_{t,k}\!=\!\log_2\!\left(\!1\!+\!\frac{ \gamma_kP_k}{1\!+\!\gamma_k\sum_{i=k+1}^{K}\!P_i}\!\right)\!, \forall k\!\in\!\left\{1,\!\cdots,\!K\!-\!1\right\},
\end{equation}
we have
$
  x\ge\frac{\left(2^{Q}-1\right)\left(1+\gamma_ky\right)}{\gamma_k}.
$
Since $2^{Q}\left(1+\gamma_ky\right)^2-\left(1+\gamma_k\left(x+y\right)\right)^2$ increases as $x$ decreases, we find that
\begin{align}\label{}
  &2^{Q}\left(1+\gamma_ky\right)^2-\left(1+\gamma_k\left(x+y\right)\right)^2 \notag \\
    \le & 2^{Q}\left(1+\gamma_ky\right)^2-\left(1+\gamma_k\left(\frac{\left(2^{Q}-1\right)\left(1+\gamma_ky\right)}{\gamma_k}+y\right)\right)^2
    \notag \\
    =&-\left(2^{Q}-1\right)2^{Q}\left(1+\gamma_ky\right)^2<0.
\end{align}
Therefore, we have $ \partial f_k\left(x,y\right)/\partial x>0$ and  $ \partial f_k\left(x,y\right)/\partial y<0$, which indicates: 1) for any given $\sum_{i=k+1}^K P_i$, the minimum $ P_k$ is obtained when the constraint \eqref{eq:P2_cons_psok} is active, and 2) the minimum $ P_k$ decreases as $\sum_{i=k+1}^K P_i$ decreases. We further find that the minimum $ P_K$ is obtained when the constraint \eqref{eq:P2_cons_psoK} is active. Then, the optimal $\bm{ P}$ can be iteratively obtained as given in  Theorem~\ref{theo:P3optalpha}. 
This completes the proof.

\section{Proof of Corollary \ref{Coro:1}}\label{app:proofCoro1}
It is easy to prove that the problem \eqref{prob:MinSumP_Simple} is feasible if \eqref{eq:coro_kg0} and \eqref{eq:coro_Kg0} can be satisfied, since $\bm{ P}^o=\left[ P^o_1,\cdots, P^o_K\right]$ would be a solution of $\bm{ P}$ to the problem \eqref{prob:MinSumP_Simple} when \eqref{eq:coro_kg0} and \eqref{eq:coro_Kg0} can be satisfied. In the following, we prove that the problem \eqref{prob:MinSumP_Simple} is infeasible when \eqref{eq:coro_kg0} and/or \eqref{eq:coro_Kg0} cannot be satisfied.
To this end, we show that~\eqref{eq:P2_cons_psok} and/or~\eqref{eq:P2_cons_psoK} cannot be satisfied  subject to \eqref{eq:P2_cons_Pk0}, when \eqref{eq:coro_kg0} and/or \eqref{eq:coro_Kg0} cannot be satisfied.

We  rewrite~\eqref{eq:P2_cons_psoK} as
\begin{equation}\label{eq:corr_P2_cons_psoK_eqv}
  2^{ Q}-1\le\left(\gamma_K-\bar{\gamma}_e\ln\left({\epsilon^{-1}}\right)2^{ Q}\right) P_K,
\end{equation}
and find that \eqref{eq:corr_P2_cons_psoK_eqv} is not achievable for any $ P_K>0$ when  \eqref{eq:coro_Kg0} is not satisfied. Thus,~\eqref{eq:P2_cons_psoK} cannot be satisfied subject to \eqref{eq:P2_cons_Pk0}, when \eqref{eq:coro_Kg0} is not satisfied.

Now, we consider the situation where
\begin{equation}\label{eq:progfeap1mids1}
  {\gamma_m\left(\!1\!-\!\bar{\gamma}_e\ln\left({\epsilon^{-1}}\right)\left(2^{ Q}\!-\!1\right)\sum_{i=m+1}^K P^o_i\right)\!-\!\bar{\gamma}_e\ln\left({\epsilon^{-1}}\right)2^{ Q}}\le0
\end{equation}
and
\begin{align}\label{eq:progfeap1mids2}
  &{\gamma_n\left(\!1\!-\!\bar{\gamma}_e\ln\left({\epsilon^{-1}}\right)\left(2^{ Q}\!-\!1\right)\sum_{i=n+1}^K P^o_i\right)\!-\!\bar{\gamma}_e\ln\left({\epsilon^{-1}}\right)2^{ Q}}>0,\notag\\
&~\forall n\in\left\{m+1,\cdots,K\right\}.
\end{align}
According to the discussion in Appendix~\ref{app:proofTheo1}, we know that
\begin{align}\label{}
\bar{\gamma}_e\ln\left({\epsilon^{-1}}\right)\le
\frac{\frac{1+\gamma_m\sum_{i=m}^{K} P_i}{1+\gamma_m\sum_{i=m+1}^{K} P_i}-2^{ Q}}
{2^{ Q}\sum_{i=m}^{K} P_i-\frac{1+\gamma_m\sum_{i=m}^{K} P_i}{1+\gamma_m\sum_{i=m+1}^{K} P_i}\sum_{i=m+1}^{K} P_i}
\end{align}
always cannot be satisfied if
\begin{align}\label{eq:proofcorrP2maopt}
&\bar{\gamma}_e\ln\left({\epsilon^{-1}}\right)\le\notag\\
&\frac{1+\frac{\gamma_m P_m}{1+\gamma_m\sum_{i=m+1}^{K} P^o_i}-2^{ Q}}
{2^{Q}\!\left(\!P_m\!\!+\!\!\sum_{i=m+1}^{K}\!P^o_i\!\right)\!\!-\!\!\left(\!1\!\!+\!\frac{\gamma_m P_m}{\!1\!+\!\gamma_m\!\sum_{i=m+1}^{K}\! P^o_i}\!\right)\!\!\sum_{i=m\!+\!1}^{K}\!P^o_i}
\end{align}
is not satisfied.
We can rewrite \eqref{eq:proofcorrP2maopt} as
\begin{align}\label{eq:corr_P2_cons_psom_eqv}
&\left(2^{ Q}\!-\!1\right)\left(\!1\!+\!\bar{\gamma}_e\ln\left({\epsilon^{-1}}\right)\sum_{i=m+1}^K P^o_i\right)\left(1\!\!+\!\gamma_m\sum_{i=m+1}^K P^o_i\right)\le \notag\\
&    \left(\!\gamma_m\!\left(\!1\!-\!\bar{\gamma}_e\ln\left({\epsilon^{-1}}\right)\left(2^{ Q}\!-\!1\right)\!\!\!\sum_{i=m+1}^K\!\! P^o_i\!\right)\!\!-\!\bar{\gamma}_e\ln\left({\epsilon^{-1}}\right)2^{ Q}\!\right)\! P_m.   \end{align}
We find that \eqref{eq:corr_P2_cons_psom_eqv} is not achievable for any $ P_m>0$ when  \eqref{eq:progfeap1mids1} and \eqref{eq:progfeap1mids2} happen. Therefore,~\eqref{eq:P2_cons_psok} cannot be satisfied subject to \eqref{eq:P2_cons_Pk0}, when \eqref{eq:coro_kg0} cannot be satisfied.
This completes the proof.

\section{Proof of Proposition~\ref{prop:feasiblePmaxmin}}\label{app:prooffeasiblePmaxmin}
We can find that the problem \eqref{prob:MaxminRs_simple} is infeasible if any of the constraints in the problem cannot be satisfied even with $\Rsmin\rightarrow0$. We can also find that the problem \eqref{prob:MaxminRs_simple} is feasible if all constraints in the problem can be satisfied with $\Rsmin\rightarrow0$.
When $\Rsmin\rightarrow0$, the constraints \eqref{eq:Pmaxmin_cons_psok} and \eqref{eq:Pmaxmin_cons_psoK}   become
\begin{align}\label{}
 & \bar{\gamma}_e\ln\left({\epsilon^{-1}}\right)\le
               \frac{\frac{1+\gamma_k\sum_{i=k}^{K} P_i}{1+\gamma_k\sum_{i=k+1}^{K} P_i}-1}
               {\sum_{i=k}^{K} P_i-\frac{1+\gamma_k\sum_{i=k}^{K} P_i}{1+\gamma_k\sum_{i=k+1}^{K} P_i}\sum_{i=k+1}^{K} P_i} =\gamma_k,   \notag\\
                &~\forall k\in\left\{1,\cdots,K-1\right\}
\end{align}
and
\begin{equation}\label{}
  \bar{\gamma}_e\ln{\left(\epsilon^{-1}\right)}\le\gamma_K,
\end{equation}
respectively.
Thus, when $\gamma_k>\bar{\gamma}_e\ln\left(\epsilon^{-1}\right)$, $\forall k\in\left\{1,\cdots,K\right\}$,  the constraints can be satisfied by having a sufficiently small $\Rsmin$. Otherwise,  \eqref{eq:Pmaxmin_cons_psok} and/or \eqref{eq:Pmaxmin_cons_psoK} cannot be satisfied even if $\Rsmin\rightarrow0$. This completes the proof.

\section{Proof of Proposition~\ref{prop:1twousersolutio}}\label{app:proofprop1}
In the proof, we first obtain the optimal $\Rsmin$,  and then get the optimal $P_1$ and $P_2$ accordingly.

The optimal $\Rsmin$, i.e., $\Rsmin^*$, is the maximum $Q$ subject to which the problem \eqref{prob:MinSumP_Simple} is feasible and the solution to the problem \eqref{prob:MinSumP_Simple} satisfies $\sum_{k=1}^KP_k^o\le P$.
According to Corollary~\ref{Coro:1}, the problem \eqref{prob:MinSumP_Simple} is feasible when both
\eqref{eq:coro_kg0} and \eqref{eq:coro_Kg0} can be satisfied.
Denote $\phi=\bar{\gamma}_e\ln\left({\epsilon^{-1}}\right)$. From \eqref{eq:coro_kg0} and \eqref{eq:coro_Kg0} with $K=2$, we have
\begin{equation}\label{eq:proofpropb1}
  2^{ Q}<\frac{\gamma_2}{\phi}=b_1
\end{equation}
and
\begin{equation}\label{eq:proofpropb2}
  2^{Q}\!\!<\!\!\frac{\phi^{-\frac{1}{2}}\!\!
  \left(4\phi^2\gamma_1\!\!-\!\!3\phi\gamma_1^2\!\!-\!6\phi\gamma_1\gamma_2\!+\!4\gamma_1^2\gamma_2\!\!+\!\!\phi\gamma_2^2\right)^{\!\frac{1}{2}}\!\!
  \!-\!\left(\!\gamma_2\!\!-\!\!\gamma_1\!\right)}{2\left(\gamma_1\!-\!\phi\right)}\!\!=\!b_2.
\end{equation}
Based on Theorem~\ref{theo:P3optalpha}, the solutions of $P_1$ and $ P_2$ to the problem \eqref{prob:MinSumP_Simple} subject to a given $ Q$ are given by
\begin{align}\label{eq:proofpropalpha01}
   P_1^o &=  \frac{\left(2^{ Q}-1\right)\left(1+\bar{\gamma}_e\ln\left({\epsilon^{-1}}\right) P^o_2\right)\left(1+\gamma_k P^o_2\right)}
     {\gamma_k\left(1-\bar{\gamma}_e\ln\left({\epsilon^{-1}}\right)\left(2^{ Q}-1\right) P^o_2\right)-\bar{\gamma}_e\ln\left({\epsilon^{-1}}\right)2^{ Q}}  \notag  \\
   &=\frac{\left(\gamma_2-\phi\right)\left(2^{ Q}-1\right)\left(\gamma_2-\gamma_1+2^{ Q}\left(\gamma_1-\phi\right)\right)}{\left(2^{ Q}\phi\!-\!\gamma_2\right)
   \left(\phi\left(\gamma_2 2^{ Q}\!+\!\gamma_1\!\left(4^{Q}\!-\!2^{ Q}\!+\!1\right)\right)\!-\!\phi^24^{ Q}-\!\gamma_1\gamma_2
   \right)}
\end{align}
and
\begin{equation}\label{eq:proofpropalpha02}
 P_2^o=\frac{2^{ Q}-1}{\gamma_2-\bar{\gamma}_e\ln\left({\epsilon^{-1}}\right)2^{Q}},
\end{equation}
respectively. Substituting \eqref{eq:proofpropalpha01} and~\eqref{eq:proofpropalpha02} into the constraint  $P_1^o+  P_2^o\le P$, we have
\begin{equation}\label{eq:proofpropb3}
  2^{ Q}\le\frac{\psi-\left(1+\phi P\right)\left(\gamma_2-\gamma_1\right)}{2\left(1+\phi P\right)\left(\gamma_1-\phi\right)}
  =b_3,
\end{equation}
where\\
\vspace{1mm}
\begin{small} $\psi\!=\!\!\sqrt{\!\left(1\!+\!\phi P\right)\!\left(4\left(1\!+\!\gamma_1 P\right)\!\left(\gamma_1\!-\!\phi\right)\left(\gamma_2\!-\!\phi\right)\!+\!\left(1\!+\!\phi P\right)\!\left(\gamma_2\!-\!\gamma_1\right)^2\right)}$.\end{small}
After performing a series of algebraic manipulations, we find that
\begin{align}\label{}
 & \text{sgn}\left\{b_1-b_3\right\}=\notag\\
 & \text{sgn}\!\left\{4\!\left(1\!+\!\phi P\right)\!\left(\gamma_1\!-\!\phi\right)\!\left(\gamma_2\!-\!\phi\right)\!\left(\left(\gamma_1\gamma_2\!-\!\phi^2\right)\!+\!\gamma_1\phi P\!\left(\!\gamma_2\!-\!\phi\right)\right)\!\right\}
\end{align}
and
\begin{equation}\label{}
  \text{sgn}\left\{b_2-b_3\right\}=\text{sgn}\left\{4\left(1+\phi P\right)\left(\gamma_1-\phi\right)^3\left(\gamma_2-\phi\right)\right\}.
\end{equation}
Since $\gamma_2>\gamma_1>\phi$ and $P>0$, we have $b_1-b_3>0$ and $b_2-b_3>0$, which indicates that  $\min\left\{b_1,b_2,b_3\right\}=b_3$. Thus, the constraint \eqref{eq:proofpropb3} is always the most stringent among the constraints \eqref{eq:proofpropb1}, \eqref{eq:proofpropb2}, and \eqref{eq:proofpropb3}.
Then, $\Rsmin^*$ in \eqref{eq:Rsstarprop1} is obtained by solving for $Q$ in the equation $2^{ Q}=b_3$.

Accordingly, $P_1^*$ in \eqref{eq:a1starprop1} and $ P_2^*$ in \eqref{eq:a2starprop1} are derived by substituting $ Q=\Rsmin^*$ into \eqref{eq:proofpropalpha01} and \eqref{eq:proofpropalpha02}, respectively. This completes the proof.

\section{Proof of Corollary~\ref{Coro:beta1increase}}\label{app:proofbeta1increase}
We find that it is difficult to directly analyze the derivative of $\beta_1^*$ with respect to $\phi$ due to the complicated expression for $\partial \beta_1^*/\partial\phi$. Thus, we rewrite $\beta_1$ in~\eqref{eq:p1optdividedbyP} as
\begin{equation}\label{}
   \beta_1^*\!=\!\frac{\rho_1\!\!+\!\!\rho_2\!\!+\!\!2\rho_1\rho_2\!\!-\!\!2\varphi^2\!\left(1\!\!+\!\!\rho_1\right)\!+\!\varphi\left(\rho_2\!-\!\rho_1\!+\!2\rho_1\rho_2\!-\!2\right)\!-\!\omega}
 {2\left(\left(1+\varphi\right)\rho_1\rho_2-\varphi^2\left(1+\rho_1\right)\right)},
\end{equation}
where $\rho_1=\gamma_1P$, $\rho_2=\gamma_2P$, $\varphi=\phi P$, and\\
\begin{small}
$\omega\!=\!\sqrt{\!\left(1\!+\!\varphi\right)\left(4\left(1\!+\!\rho_1\right)\left(\rho_1\!-\!\varphi\right)\left(\rho_2-\varphi\right)+\left(1+\varphi\right)\left(\rho_2-\rho_1\right)^2\right)}$.
\end{small}
For any given $P>0$, taking the  derivative of $\beta_1^*$ with respect to $\varphi$, we have
\begin{equation}\label{}
  \frac{\partial \beta_1^*}{\partial\varphi}\!=\!\frac{4\!\left(1\!
\!+\!\!\varphi\right)\!\left(\rho_1\!\!-\!\!\varphi\right)\!\left(\rho_2\!\!-\!\!\varphi\right)\!\left(1\!\!+\!\!\rho_1\right)\!\left(\!2\varphi\!\!+\!\!\varphi^2\!\!-\!\!\rho_1\!\!-\!\!\rho_2\!\!-\!\!\rho_1\rho_2\!\!+\!\!\omega\!\right)}
  {\omega\left(\left(1+\varphi\right)\left(\varphi\left(\rho_1+\rho_2\right)-2\rho_1\rho_2\right)+\varphi\omega\right)^2}\!.
\end{equation}
After performing a series of algebraic manipulations, we find that
\begin{equation}\label{}
   \text{sgn}\left\{\frac{\partial \beta_1^*}{\partial\varphi}\right\}= \text{sgn}\left\{\varphi^2-2\varphi\rho_1+\rho_1\rho_2+\left(2+\varphi\right)\left(\rho_2-\rho_1\right)\right\}.
\end{equation}
Note that $\rho_2\ge\rho_1>\varphi$, since $P>0$ and $\gamma_2\ge\gamma_1>\phi$. When $\rho_2\ge\rho_1>\varphi$, we find that
\begin{align}\label{}
  &\varphi^2-2\varphi\rho_1+\rho_1\rho_2+\left(2+\varphi\right)\left(\rho_2-\rho_1\right)>\notag\\
  &\left(\rho_1-\varphi\right)^2+\left(2+\varphi\right)\left(\rho_2-\rho_1\right)>
  0.
\end{align}
 Thus, $\partial \beta_1^*/\partial\phi>0$.
This completes the proof.

\section{Proof of Proposition~\ref{prop:propNOMAOMAc}}\label{app:proofpropNOMAOMAc}
\begin{figure}[!t]
\centering
\includegraphics[width=0.9\columnwidth]{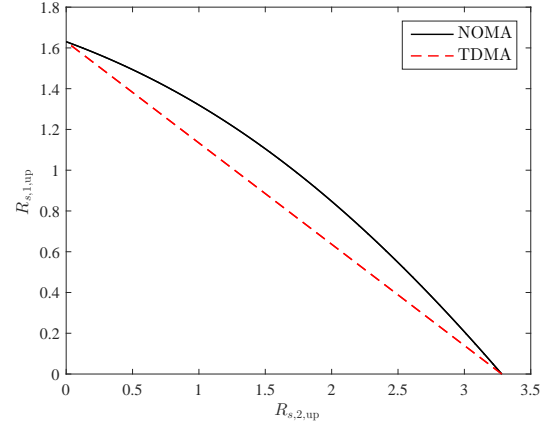}
\caption{An illustration of the boundary of confidential information rate pairs with $\gamma_1\neq\gamma_2$.}
\vspace{0mm}  \label{fig:R2VsR1}
\end{figure}

Denote the achievable confidential information rate pairs of user 1 and user 2 by the TDMA scheme and NOMA scheme as $\left(R_{s,1}^{\TDMA},R_{s,2}^{\TDMA}\right)$ and $\left(R_{s,1}^{\NOMA},R_{s,2}^{\NOMA}\right)$, respectively.
We know that the boundary of rate pairs is obtained when the available resources are all used, i.e., $t_1+t_2=1$ for the TDMA scheme and $ P_1+ P_2=P$ for the NOMA scheme. Thus, based on \eqref{eq:OMARsk}, the upper boundary of the rate pairs achieved by the TDMA scheme is given by
\begin{align}\label{eq:boundary_OMApair}
  &\left(R_{s,1,\up}^{\TDMA}=(1-t_2)\log_2\left(\frac{1+\gamma_1P}{1+\bar{\gamma}_e\ln\left(\epsilon^{-1}\right)P}\right),\right.\notag\\
   &~~\left.R_{s,2,\up}^{\TDMA}=t_2\log_2\left(\frac{1+\gamma_2P}{1+\bar{\gamma}_e\ln\left(\epsilon^{-1}\right)P}\right)\right).
\end{align}
From \eqref{eq:Pmaxmin_cons_psok} and \eqref{eq:Pmaxmin_cons_psoK}, the upper boundary of the rate pairs achieved by the NOMA scheme is given by
\begin{align}\label{eq:boundary_NOMApair}
  &\left(R_{s,1,\up}^{\NOMA}=\log_2\left(\frac{\left(1+\gamma_1P\right)\left(1+ \bar{\gamma}_e\ln\left(\epsilon^{-1}\right)P_2\right)}{\left(1+ \gamma_1P_2\right)\left(1+\bar{\gamma}_e\ln\left(\epsilon^{-1}\right)P\right)}\right),\right.\notag\\
  &~~\left.
  R_{s,2,\up}^{\NOMA}=\log_2\left(\frac{1+ \gamma_2P_2}{1+ \bar{\gamma}_e\ln\left(\epsilon^{-1}\right)P_2}\right)
  \right).
\end{align}
From \eqref{eq:boundary_OMApair} and \eqref{eq:boundary_NOMApair}, we find that
\begin{align}\label{}
  &\left(R_{s,1,\up}^{\TDMA}\left(t_2\!=\!0\right),R_{s,2,\up}^{\TDMA}\left(t_2=0\right)\right)\notag\\
  =&\left(R_{s,1,\up}^{\NOMA}\left( P_2=0\right),R_{s,2,\up}^{\NOMA}\left( P_2=0\right)\right)
\end{align}
and
\begin{align}\label{}
  &\left(R_{s,1,\up}^{\TDMA}\left(t_2=1\right),R_{s,2,\up}^{\TDMA}\left(t_2=1\right)\right)\notag\\
  =&\left(R_{s,1,\up}^{\NOMA}\left( P_2=P\right),R_{s,2,\up}^{\NOMA}\left( P_2=P\right)\right).
\end{align}
We also find that the rate pairs achieved by the TDMA scheme are in a convex region with the affine boundary.
Thus, we can prove that any rate pair achieved by the TDMA is strictly smaller than a rate pair that can be achieved by the NOMA scheme if the upper boundary of the rate pairs achieved by the NOMA scheme is a concave function. We can also prove that any rate pair achieved by the TDMA is the same to a rate pair achieved by the NOMA scheme if the upper boundary of the rate pairs achieved by the NOMA scheme is a affine function. An illustration of the boundary of rate pair is given in Figure~\ref{fig:R2VsR1}.

In the following, we investigate the upper boundary of the rate paired achieved by the NOMA scheme, i.e., $\left(R_{s,1,\up}^{\NOMA},R_{s,2,\up}^{\NOMA}\right)$.
From~\eqref{eq:boundary_NOMApair}, we can rewrite $R_{s,1,\up}^{\NOMA}$ as a function of $R_{s,2,\up}^{\NOMA}$, which is given by
\begin{equation}\label{}
  R_{s,1,\up}^{\NOMA}=\log_2\left(\frac{\left(1+\rho_1\right)\left(\rho_2-\varphi\right)}
  {\left(1+\varphi\right)\left(\rho_2-\rho_1+\left(\rho_1-\varphi\right)2^{ R_{s,2,\up}^{\NOMA}}\right)}\right).
\end{equation}
where $\rho_1=\gamma_1P$, $\rho_2=\gamma_2P$, and $\varphi=\bar{\gamma}_e\ln\left(\epsilon^{-1}\right)P$.
Taking the second order derivative of  $R_{s,1,\up}^{\NOMA}$ with respect to $R_{s,2,\up}^{\NOMA}$, we have
\begin{equation}\label{eq:deriv2Rs1NOMA2Rs2NOMA}
  \frac{\partial^2 R_{s,1,\up}^{\NOMA}}{\partial \left(R_{s,2,\up}^{\NOMA}\right)^2}=-\frac{2^{R_{s,2,\up}^{\NOMA}}\left(\ln(2)\right)^2
  \left(\rho_2-\rho_1\right)\left(\rho_1-\varphi\right)}
  {\left(\rho_2-\rho_1+\left(\rho_1-\varphi\right)2^{ R_{s,2,\up}^{\NOMA}}\right)^2}.
\end{equation}
From \eqref{eq:deriv2Rs1NOMA2Rs2NOMA}, we see that  $ \partial^2 R_{s,1,\up}^{\NOMA}/
\partial \left(R_{s,2,\up}^{\NOMA}\right)^2<0$ if $\rho_1<\rho_2$, i.e., $\gamma_1<\gamma_2$. Thus, the  upper boundary of the rate pairs achieved by the NOMA scheme is a concave function when $\gamma_1\neq\gamma_2$. When $\rho_1=\rho_2$, i.e., $\gamma_1=\gamma_2$, we have
\begin{equation}\label{}
  R_{s,1,\up}^{\NOMA}=\log_2\left(\frac{1+\rho_1}{1+\varphi}\right)-R_{s,2,\up}^{\NOMA}.
\end{equation}
Thus, the  upper boundary of the rate pairs achieved by the NOMA scheme is an affine function when $\gamma_1=\gamma_2$. This completes the proof.

\balance


\end{document}